\documentclass{SCGE}
\usepackage{soul,wasysym,amssymb,graphicx,bm}
\usepackage[colorlinks,linkcolor=blue,citecolor=blue,urlcolor=blue]{hyperref}

\let\citedash\relax
\makeatletter \providecommand{\citedash}{\hbox{-}\penalty\@m}
\makeatother

\begin{document}
\begin{picture}(0,0){\rm
\put(0,-20){\makebox[160truemm][l]{\bf {\sanhao\raisebox{2pt}{.}}
Article  {\sanhao\raisebox{1.5pt}{.}}}}}
\put(0,-34){\jiuwuhao {\textcolor[rgb]{0.5,0.5,0.5}{\sf 
}}}
\end{picture}

\def\bm{\boldsymbol}

\def\dl{\displaystyle}
\def\du{\end{document}}
\def\d{{\rm d}}
\def\e{{\rm e}}
\def\i{{\rm i}}

\Year{2016} %
\Month{January} %
\Vol{00} %
\No{0} %
\BeginPage{0} %
\AuthorMark{{\rm L. Shao and N. Wex}}  %
\DOI{} %
\ArtNo{000000}

\title[Tests of Gravitational Symmetries with Radio Pulsars]{Tests of
Gravitational Symmetries with Radio Pulsars} 

\author{Lijing Shao$^{1,2}$ and Norbert Wex$^{3}$}{}

\address[{\rm1}]{Max Planck Institute for Gravitational Physics (Albert Einstein
Institute), Am M\"uhlenberg 1, D-14476 Potsdam-Golm, Germany}
\address[{\rm2}]{School of Physics and State Key Laboratory of Nuclear Physics
and Technology, Peking University, Beijing 100871, China}
\address[{\rm3}]{Max-Planck-Institut f\"ur Radioastronomie, Auf dem H\"ugel 69,
D-53121 Bonn, Germany}

\maketitle \vspace{-3.5mm}{\footnotesize\begin{center} Received January 1, 2016;
  accepted January 1, 2016; published online January 1, 2016
\end{center}}\vspace*{-5mm}

\begin{center}
\rule{16.5cm}{0.4pt}
\parbox{16.5cm}
{\begin{abstract} 
  Symmetries play important roles in modern theories of physical laws. In this
  paper, we review several experimental tests of important symmetries associated
  with the gravitational interaction, including the universality of free fall
  for self-gravitating bodies, time-shift symmetry in the gravitational
  constant, local position invariance and local Lorentz invariance of gravity,
  and spacetime translational symmetries. Recent experimental explorations for
  post-Newtonian gravity are discussed, of which, those from pulsar astronomy
  are highlighted. All of these tests, of very different aspects of gravity
  theories, at very different length scales, favor to very high precision the
  predictions of the strong equivalence principle (SEP) and, in particular,
  general relativity which embodies SEP completely. As the founding principles
  of gravity, these symmetries are motivated to be promoted to even stricter
  tests in future.
\end{abstract}}
\end{center}\vspace*{-0.6cm}

\begin{center}
\parbox{16.5cm}
{\bf\jiuhao Gravitation, Radio Pulsars, Spacetime Symmetry}%
\end{center}

\begin{center}
{}%
\CITA    
\end{center}

\textwidth=178truemm \textheight=236truemm%

\wuhao\vspace*{1.5mm}

\begin{multicols}{2}

\renewcommand{\baselinestretch}{1.08} \baselineskip 12.2pt\parindent=10.8pt


\section{Introduction}
\label{sec:intro}

One hundred years ago, in November 1915, Albert Einstein presented the final
form of his field equations of gravitation (without cosmological term) to the
Prussian Academy of Science \cite{ein15}. With this publication, general
relativity (GR) was finally completed as a logically consistent physical theory.
Already one week before, based on the vacuum form of his field equations,
Einstein was able to demonstrate that GR naturally explains the anomalous
advance of the Mercury perihelion\cite{ein15a}. Since then, GR was confronted
with various dedicated  ground- and space-based precision
experiments~\cite{will14}. GR has passed all these tests with flying colors,
including the, since then much improved, initial three classical tests, namely
the perihelion advance of Mercury, the deflection of light by the Sun, and the
gravitational redshift. Various, quite diverse  aspects of gravity are covered
by GR: the (accelerated) expansion of the Universe, the reaction of the
gravitational radiation of inspiralling binaries, the existence of black holes
as the end products of gravitational collapse, the dragging of spacetime by a
massive rotating body, the quantum phase evolution of cold atoms in the
gravitational field, just to name a few. It is astonishing that a single gravity
theory with only two parameters (namely the gravitational constant $G$ and the
cosmological constant $\Lambda$) is able to interpret all of these apparently
vastly unrelated phenomena, to very high precision, at different length scales. 

Nevertheless, there are various sophisticated reasons to search for gravity
theories beyond Einstein's GR. Some of the most important ones are listed below.  
\begin{itemize} 
  \item Einstein's GR does not employ quantum principles. Quantum nature of
    microscopic objects was well established to exquisite precision, and has
    already been integrated into our everyday life as a necessity. Quantum field
    theories that implement basic quantum ingredients describe all forces other
    than the gravity (namely, the electromagnetic force, the weak force, and the
    strong force), and they are tested against numerous data from particle
    accelerators, with many important predictions verified highly
    satisfactorily~\cite{pdg14}. However, GR is known to
    have difficulties to be renormalized, which is one of the basic requirements
    for a modern quantum field theory.
  \item A related problem is about the appearance of singularities in spacetime
    in GR, under rather general circumstances. The singularity theorems of GR
    show that such singularities are inevitable in many physical situations
    \cite{he73}. There is hope that, by incorporating quantum fluctuations to
    GR, these singularity problems will be cured~\cite{ame13}.  
  \item During recent decades, new phenomena from astrophysics and cosmology ask
    for dark ingredients in our Universe (namely, dark matter and dark energy).
    Their gravitational behaviors are crucial in explaining continuously
    accumulating new observations. While these dark ingredients have not (yet)
    shown up as new particles in dedicated experiments in ground-based
    laboratories, some theories suggest to modify GR to avoid the need of dark
    matter and/or dark energy~\cite{cfps12}.  
  \item Besides the aforementioned dark matter and dark energy, cosmological
    data are favoring an exponentially growing inflation era during the very
    early Universe.  Such a scenario needs extra inputs beyond GR for the
    dynamics of spacetime, for example, one or several scalar inflatons that
    drive the exponential expansion. This might suggests a scalar degree in
    gravity, in addition to the canonical rank-2 metric tensor in
    GR~\cite{cfps12}.  
\end{itemize}

Although an ultimate quantum gravity theory is still under construction,
different modified gravity theories were built tentatively to account for one or
several of the aforementioned problems. Scalar-tensor theories of gravity have
been studied since the 1940s, with Jordan-Fierz-Brans-Dicke gravity as their
most prominent representative (see Ref.~\cite{goe12}, and references therein).
An important motivation behind Jordan-Fierz-Brans-Dicke gravity was the
promotion of $G$ to a dynamical field, by this introducing a time-varying
gravitational constant in an expanding Universe \cite{jor55}. 

More elaborate gravity theories have been ``designed'' to address (at least
parts of) the problem of the dark sector. One of the early attempts was
Bekenstein's tensor-vector-scalar (TeVeS) theory, a covariant field theory in
the modified Newtonian dynamics paradigm, aiming to explain dark matter as a
manifest of a modification to GR~\cite{bek04,bek04e}. 

Ho{\v r}ava suggested different scaling indices for temporal and spatial
components of a fossilized spacetime (that has a preferred time direction) to
achieve a power-counting renormalizable, UV complete, gravity theory
\cite{hor09}.  These new gravity theories were refined later by other authors to
evade tensions with existing and upcoming experiments, and cure possible
intrinsic problems in the theory \cite{bps11}. 

As a general rule, new theories intend to break some foundational symmetries in
GR. We are not to state that breaking these symmetries is necessarily bad,
instead, these possibilities should be explored as clues to
alternatives~\cite{bbc+15}. On top of that, this generally introduces a
parameter space, in which GR can be tested \cite{will93,dam09}. Nevertheless,
from pure aesthetic viewpoint, a theory with less symmetry could be less
appealing. Extending GR with quantum principles is widely believed to require
deep insights into GR's founding principles and possibly do some
modification/generalization/extension to them.  Ultimately, one should resort to
discriminating experiments to judge between different choices of theories.
Therefore, experimentally exploring the foundational symmetries in GR is highly
desired.

Speaking of foundational symmetries, the {\em strong equivalence principle}
(SEP) represents a highly condensed wisdom for GR \cite{mtw73}.  Based on early
work by Robert Dicke, Clifford Will nicely constructed and popularized SEP with
three ingredients~\cite{will93,will14} (see also the discussion in
Ref.~\cite{dls15}): 
\begin{itemize} 
  \item Validity of the weak equivalence principle for test particles (WEP) is
    extended to self-gravitating bodies (GWEP).  It means that in the external
    gravitational field, bodies of negligible or non-negligible self-gravity
    should fall at the same rate, regardless of the detailed composition (for
    example, with different number of quarks and electrons, different fraction
    $\varepsilon$ of gravitational binding energy, et cetera). It is a
    nontrivial extension of Galileo's universality of free fall to include also
    bodies where the gravitational binding contributes a significant fraction to
    their mass, e.g.\  the Sun ($\varepsilon \sim 10^{-6}$) or neutron stars
    ($\varepsilon \sim 0.1$).  
  \item Local Lorentz invariance in gravity. It states that, no matter of the
    velocity of experimental apparatus, the outcome of local experiments,
    gravitational or non-gravitational, should be described by the same set of
    physical laws.  
  \item Local position invariance in gravity. It states that, no matter of when
    and where the local experiments are performed, the outcome should be
    interpreted by the same set of physical laws.
\end{itemize}

The SEP describes the general rules for the outcome of experiments, both
gravitational and non-gravitational. It is indeed lying to the heart of GR.
Actually, there are good arguments that GR is the only valid gravity that
respects SEP in its entirety \cite{will14}. Therefore, probing the building
blocks of SEP probes the deep foundational principles/symmetries of GR.

In this review, we will touch on some experimental tests on different aspects of
SEP.  Precision pulsar timing experiments are highlighted, because they not only
provide the most limiting constraints in most areas, but also probe certain
aspects of strong-field gravity, in particular those associated with neutron
stars, the supposedly most compact material bodies in nature \cite{sta03,wex14}.
In the next section, the universality of free fall for self-gravitating bodies
is reviewed, with a focus on experiments that constrain the violation of GWEP in
the quasi-stationary strong-field regime. Closely related to a violation of GWEP
is the existence of dipolar radiation, in particular if the modification of GR
is associated with additional ``gravitational charges''. This is a powerful
discriminant in asserting the principle from orbital decays of binary pulsars,
capable of capturing deviations from GR that evade free-fall tests. In
Section~\ref{sec:lpi} and Section~\ref{sec:lli}, the local position invariance
and local Lorentz invariance in gravity are examined in two generic, overlapping
but not equivalent, frameworks for post-Newtonian gravity, namely, the
parametrized post-Newtonian (PPN) formalism~\cite{will14} and the standard-model
extension (SME)~\cite{kos04}.  Again, pulsar timing turns out one of the best
testbeds for any tiny deviations from GR permitted by these two frameworks. In
Section~\ref{sec:trans}, we review results on energy-momentum conservation laws
in post-Newtonian gravity in the PPN formalism~\cite{will93}.  Finally, in
Section~\ref{sec:Gdot} we briefly look into aspects of a time-varying
gravitational constant, whereby we underline the complementary aspect of pulsar
tests. Section~\ref{sec:sum} summarizes the paper, and gives a short discussion
on possible strong-field effects and the upcoming gravitational-wave
experiments~\cite{ss09,bs14}. 

\section{Universality of Free Fall for Self-Gravitating Bodies}
\label{sec:uff}

As mentioned in Section~\ref{sec:intro}, SEP extends WEP to the gravitational
WEP (GWEP), i.e.~the universality of free fall for self-gravitating bodies. In
GR, GWEP is fulfilled, i.e.\ in GR the world line of a body is independent of
its chemical composition and gravitational binding energy. Therefore, a
detection of a violation of GWEP, one of the three pillars of 
SEP (see Sec.~\ref{sec:intro}), would directly falsify GR. On the other hand,
alternative theories of gravity generally violate GWEP. This is also the case
for most metric theories of gravity, which by definition fulfill WEP
\cite{will93}. For a weakly self-gravitating body in a weak external
gravitational field one can simply express a violation of GWEP as a difference
between inertial (I) and (passive) gravitational (G) mass that is proportional
to the gravitational binding energy, $E_{\rm grav}$, of the body,
\begin{equation}\label{eq:mGmI} 
  \frac{m_{\rm G}}{m_{\rm I}} \simeq 
    1 + \eta_{\rm N} \, \frac{E_{\rm grav}}{m_{\rm I} c^2} = 
    1 + \eta_{\rm N} \, \varepsilon \;.
\end{equation}
The Nordtvedt parameter $\eta_{\rm N}$ is a theory dependent constant. In the
parameterized post-Newtonian (PPN) framework, $\eta_{\rm N}$ is given as a
combination of different PPN parameters (see Ref.~\cite{will93} for details). As
a consequence of Eq.~(\ref{eq:mGmI}), the Earth ($\varepsilon \approx -5 \times
10^{-10}$) and the Moon ($\varepsilon \approx -2 \times 10^{-11}$) would fall
differently in the gravitational field of the Sun (Nordtvedt effect
\cite{nor68}). The parameter $\eta_{\rm N}$ is therefore tightly constrained by
the lunar-laser-ranging (LLR) experiments to $\eta_{\rm N} = (3.0 \pm 3.6)
\times 10^{-4}$, which is in perfect agreement with $\eta_{\rm N} = 0$, the
prediction of GR \cite{mhb12}.

In view of the smallness of the fractional self-gravity of Earth and Moon, the
LLR experiment says nothing about strong-field aspects of GWEP. GWEP could still
be violated for extremely compact objects, like neutron stars ($\varepsilon \sim
0.1$), meaning that a neutron star would feel a quite different acceleration in
an external gravitational field than a weakly self-gravitating body. Since
beyond the first post-Newtonian approximation there is no general PPN formalism
available, and Eq.~(\ref{eq:mGmI}) is not applicable for strongly
self-gravitating masses, discussions of GWEP violation in this regime are best
done within theory-specific frameworks. A particularly suitable framework that
nicely allows the study of various strong-field deviations from GR, is the
two-parameter class of mono-scalar-tensor theories $T_1(\alpha_0,\beta_0)$ of
Refs.~\cite{de93,de96a}, where the scalar field is sourced by the trace of the
energy-momentum tensor. In this extension of the Jordan-Fierz-Brans-Dicke
theory, the coupling strength is a (linear) function of the scalar field,
$\alpha(\varphi) = \alpha_0 + \beta_0 (\varphi - \varphi_0)$, with $\varphi_0$
being the asymptotic value of the scalar field $\varphi$.
Jordan-Fierz-Brans-Dicke theory is recovered for $\beta_0 = 0$. As discovered by
Damour and Esposito-Far{\`e}se, in such theories, for certain (negative) values
of $\beta_0$, one finds significant strong-field deviations from GR, and a
correspondingly strong violation of GWEP for neutron stars. To illustrate this
violation of GWEP, it is sufficient to look at the leading order ``Newtonian''
terms in the equations of motion of a three body system, with masses $m_A$ at
(coordinate) locations ${\bf r}_A$ ($A = 1,2,3$) \cite{de92},
\begin{equation}\label{eq:3bodyEOM}
  \ddot{\bf r}_A = -\sum_{B \ne A} {\cal G}_{AB} m_B 
                   \frac{{\bf r}_A - {\bf r}_B}{|{\bf r}_A - {\bf r}_B|^3} \;,
\end{equation}
where the body-dependent effective gravitational constant ${\cal G}_{AB}$ is
related to the Newtonian gravitational constant $G$ (as measured in a
Cavendish-type experiment) by
\begin{equation}\label{eq:Geff}
  {\cal G}_{AB} = G \, \frac{1 + \sigma_A\sigma_B}{1 + \alpha_0^2} \;.
\end{equation}
The quantity $\sigma_A$ denotes the effective scalar coupling of mass $m_A$,
which gives the total scalar charge of body $A$ by $\omega_A = -\sigma_A m_A$.
For weakly self-gravitating bodies $\sigma_A \simeq \alpha_0$. $\sigma_A$ is a
body-dependent quantity, in general different for masses with different
gravitational binding energy. In other words, the gravitational interaction
depends on the internal structure of the bodies, and therefore GEWP is violated.
In GR on the other hand, one can introduce one effective mass, due to the
effacement of the internal structure \cite{dam87}, leading to a universality of
free fall for self-gravitating bodies.

Eq.~(\ref{eq:3bodyEOM}) implies that, two components of a binary system, say
$m_1$ and $m_2$, will fall differently in the external field of a distant
($|{\bf r}_3| \gg |{\bf r}_1|,|{\bf r}_2|$) third body $m_3$, since the
accelerations caused by the external mass are 
\begin{equation}
  \ddot{\bf r}_1^{\rm ext} \simeq \frac{1 + \sigma_1\sigma_3}{1 + \alpha_0^2}\, 
                                  {\bf g}^{\rm ext}
  \quad \mbox{and} \quad
  \ddot{\bf r}_2^{\rm ext} \simeq \frac{1 + \sigma_2\sigma_3}{1 + \alpha_0^2}\, 
                                  {\bf g}^{\rm ext} \,,
\end{equation}
respectively, where ${\bf g}^{\rm ext} \equiv Gm_3{\bf r}_3/r_3^3$. The
violation of GWEP can therefore be written as 
\begin{equation}\label{eq:DeltaGWEP}
  \Delta \cdot {\bf g}^{\rm ext} \equiv
  \ddot{\bf r}_1^{\rm ext} - \ddot{\bf r}_2^{\rm ext}
  \simeq (\sigma_1 - \sigma_2)\sigma_3 \cdot {\bf g}^{\rm ext} \;.
\end{equation}
In the final step, we have used the fact that, Solar system experiments already
imply $\alpha_0^2 \ll 1$. 

\begin{figure}[H]
  \centering
  \includegraphics[width=9cm]{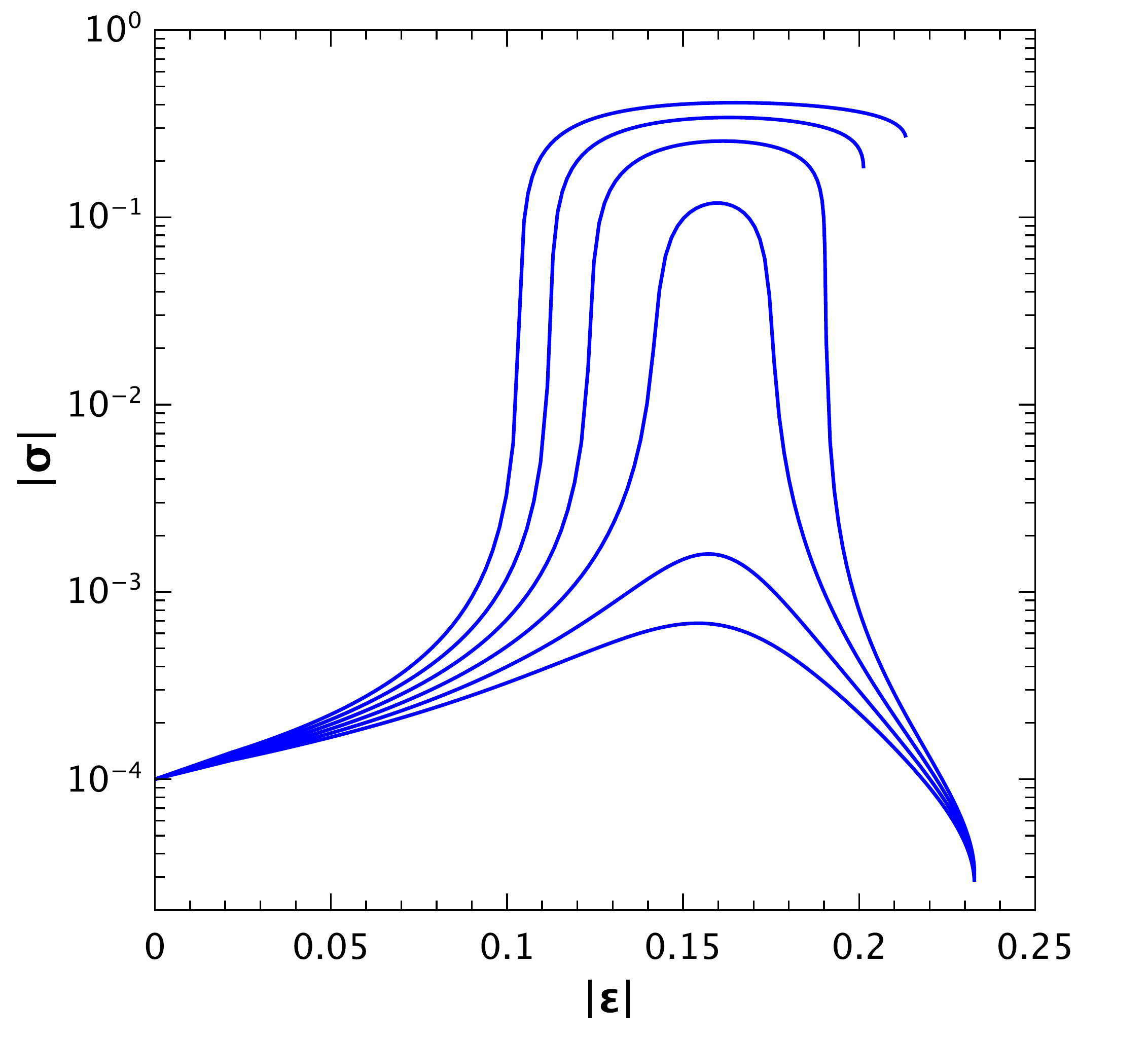}
  \caption{Effective scalar coupling $\sigma$ as a function of the fractional
    binding energy $\varepsilon$ of a (stable) neutron star, for
    $T_1(\alpha_0,\beta_0)$ scalar-tensor theories of gravity. To calculate the
    blue lines, $\alpha_0 = 10^{-4}$ and $\beta_0 = -4.0$ (bottom), $-4.2$,
    $-4.4$, $-4.6$, $-4.8$, and $-5.0$ (top) was used. Furthermore, equation of
    state AP4 (as in Ref.~\cite{lp01}) was assumed, when calculating the
    structure of the neutron stars. As one can see, if $\beta_0$ is not too
    negative, then neutron stars tend to descalarize when approaching their
    maximum mass (cf.\ Ref.~\cite{stob14}).}
  \label{fig:sig} 
\end{figure}

For neutron stars, depending on $\beta_0$ and their masses, $\sigma_A$ can
deviate significantly from $\alpha_0$. In fact, even for $\alpha_0 = 0$,
$\sigma_A$ can in principle be of order unity due to spontaneous scalarization
\cite{de93}.  Figure~\ref{fig:sig} shows $\sigma_A$ as a function of the
fractional gravitational binding energy of a neutron star for specific values of
$\beta_0$.  From this, one clearly sees the strong non-linearity of the
violation of GWEP in the strong-field regime of these specific theories of
gravity. 

For such strong-field aspects of GWEP, the best current tests do come from the
timing observations of millisecond pulsars in wide orbits with white dwarf
companions. If there is a violation of GWEP associated with the strong internal
gravity of neutron stars, then the gravitational field of the Milky Way at the
location of the pulsar (typically $|{\bf g}^{\rm ext}| \approx 2 \times
10^{-8}\,{\rm cm}\,{\rm s}^{-2}$) would polarize the binary orbit in a
characteristic way \cite{ds91}, causing, most importantly, a gradual periodic
change of the orbital eccentricity (gravitational Stark effect). Currently, the
best limits on $\Delta$ for neutron stars are obtained from the so-called
Damour-Sch\"afer test, which is based on probabilistic considerations for
small-eccentricity binary pulsars \cite{ds91}. In such binary systems, the
observed eccentricity is a combination of the forced eccentricity ${\bf e}_{\rm
F}$ (in direction of the projected ${\bf g}^{\rm ext}$) and an intrinsic
eccentricity ${\bf e}_{\rm R}$, that has a fixed length and rotates in the
orbital plane with the frequency of the relativistic periastron advance
$\dot\omega$ (analogue to Figure~\ref{fig:ecc:a1} below). For a binary, with
orbital frequency $n_b$ and semi-major axis $a$ one finds 
\begin{equation}\label{eq:eF_GWEP}
   {\bf e}_{\rm F} = \frac{3\Delta}{2 a n_b \dot\omega} \, 
                     {\bf g}^{\rm ext}_\perp \,.
\end{equation}
Consequently, the wider the orbit the stronger the polarizing effect would be.
Note however, if $\dot\omega$ becomes comparable to the rotation of the system
in the Galaxy, then Eq.~(\ref{eq:eF_GWEP}) can no longer be applied, as the
underlying assumptions for its derivation break down. 

By statistically excluding (near) alignment between ${\bf e}_{\rm F}$ and ${\bf
e}_{\rm R}$, one can generically place limits on $|{\bf e}_{\rm F}|$, and
therefore on $\Delta$, without knowledge of the actual length of ${\bf e}_{\rm
R}$. Combining a whole population of pulsar -- white dwarf systems, Stairs et al.
\cite{sfl+05} have obtained the so far best limit for a violation of GWEP by a
strongly self-gravitating body,\footnote{Note, the limit in Ref.~\cite{gsf+11}
is flawed, as explained in Ref.~\cite{wex14}.}
\begin{equation}
  |\Delta| < 0.0056 \quad \mbox{(95\% CL)} \;.
\end{equation}
In view of Eq.~(\ref{eq:DeltaGWEP}), this limit however is not very
constraining. For the white dwarf and the Galactic field $\sigma_2 \simeq
\sigma_3 \simeq \alpha_0$, resulting in a weak limit on $|\sigma_1 -
\alpha_0|$ for a neutron star since $|\alpha_0|\ll 1$. 

There is an underlying assumption in the Damour-Sch\"afer test when combining
multiple systems, which is related to the mass dependence of a GWEP violation.
Constraining a $\Delta$ from a set of pulsar -- white dwarf systems in a generic
way requires the assumption that $\Delta$ is practically independent of the mass
of the neutron star, as these systems have different pulsar masses. Even in the
absence of non-perturbative behavior, one finds rather large deviations from
that assumption along the range of observed neutron star masses. In the
presence of non-perturbative strong-field effects, like the spontaneous
scalarization mentioned above (see Figure~\ref{fig:sig}), this assumption is not
applicable at all. For this reason it is desirable to have direct tests, i.e.
tests based on long term timing observations of individual systems with well
known masses, where one directly constrains $\dot e$ (see Ref.~\cite{fkw12} for
details). As it requires a number of conditions to be met, like high timing
precision and knowledge of the orbital orientation, only few systems turn out to
be suitable at present. In Ref.~\cite{fkw12} two binary pulsar systems have been
identified as particularly suitable for a direct test of a GWEP violation,
PSRs~J1713+0747 and J1903+0327. While the work on PSR~J1713+0747 is still in
progress, preliminary results for PSR~J1903+0327 have been published in
Ref.~\cite{fkw12}, giving a rather weak constraint of $|\Delta| < 0.1$. Although
this test will improve with time, through continuous timing, it will soon become
mostly obsolete, because of a recent discovery of PSR~J0337+1715, which is a
member of a hierarchical triple system \cite{rsa+14}. In this system we have a
1.44\,$M_\odot$ pulsar in a 1.63-day orbit with a 0.2\,$M_\odot$ white dwarf.
This inner binary is in a 327-day orbit with a 0.4\,$M_\odot$ white dwarf, and
consequently experiences an external acceleration $|{\bf g}^{\rm ext}| \sim
0.2\,{\rm cm}\,{\rm s}^{-2}$, which is seven orders of magnitude larger than the
one of the Galactic gravitational field given above. Consequently,
PSR~J0337+1715 might soon allow for a significantly improved limit on $\Delta$.
As already argued in Ref.~\cite{fkw12}, pulsars in hierarchical triple systems
would be the ideal laboratories for testing GWEP, as they combine a strong
external field ${\bf g}^{\rm ext}$  with a large fractional binding energy 
$\varepsilon$. In fact, simulations in Ref.~\cite{shao16} indicate 
that a limit of $|\Delta| \lesssim 10^{-7}$ might soon be achievable.
Based on mock data simulations, Berti
et al.\ \cite{bbc+15} have demonstrated the potential of PSR~J0337+1715 to
constrain scalar-tensor theories, in particular with the timing capabilities of
future radio telescopes, like the Square Kilometre Array (SKA).

There are, however, limitations to what in terms of GWEP violation can be tested
with PSR~J0337+1715. For instance, if gravity only deviates (significantly) from
GR in the strong field regime, and hence the white dwarfs do not develop any
relevant additional gravitational charges, then PSR~J0337+1715 does not provide
any (relevant) constraints. To illustrate this, let us look at those
$T_1(\alpha_0,\beta_0)$ theories that exhibit spontaneous scalarization in the
strong gravity of a neutron star. In such a case we find for the two white
dwarfs $\sigma_2 = \sigma_3 = 0$, which leads to ${\cal G}_{12} = {\cal G}_{13}
= {\cal G}_{23} = G$, no matter how strongly the pulsar ($m_1$) is scalarized
(see discussion in Ref.~\cite{fkw12}). Since $\sigma_3 = 0$, one finds $\Delta =
0$ from Eq.~(\ref{eq:DeltaGWEP}). An obvious other ``blind-spot'' of such a GWEP
test is the situation where $\sigma_1 \simeq \sigma_2$ (cf.\
Eq.~(\ref{eq:DeltaGWEP})).  Indeed, in $T_1(\alpha_0,\beta_0)$ there is always a
value for $\beta_0$, depending on the mass of the neutron star, where $\sigma_1
= \sigma_2$. Another example is Bekenstein's TeVeS theory~\cite{bek04}, where
neutron stars and white dwarfs do have the same effective scalar coupling
\cite{fwe+12}. 

Fortunately, there are other consequences of a violation of SEP, which are
closely related to a violation of the universality of free fall of
self-gravitating masses, and can be used as a test to overcome these gaps. One
is the modification of the moment-of-inertia of the spinning pulsar, as it moves
on an eccentric orbit around its companion. Such a consequence has generically
been well constrained with the help of the double pulsar \cite{kw09}. The other
consequence of a violation of GWEP is a modification of the gravitational-wave
damping in a binary pulsar system. In general, these modifications introduce a
dipolar component, affecting the equations of motion already at the 1.5
post-Newtonian level, i.e.\ $(c/v_{\rm orb})^2$ times higher than the quadrupolar
contribution in GR, where $v_{\rm orb}$ is the characteristic velocity of binary
motion~\cite{will93}. But even in the absence of a dipolar contribution, there
is still a modification at the 2.5 post-Newtonian level (see Ref.~\cite{mw13}
for the equations of motion in scalar-tensor theories). The most important
consequence in binary pulsars, related to these modifications, is a change in
the orbital period decay rate, $\dot{P}_b$. In particular systems which show a
high asymmetry in compactness should be affected, as there one generally would
expect a dominating dipolar contribution. Within the $T_1(\alpha_0,\beta_0)$
class of gravity theories, to give a concrete example, one finds
\begin{equation}\label{eq:PbdotD}
  \dot{P}_b \simeq -\frac{4\pi^2}{P_b} \, \frac{G}{c^3} \, \frac{m_1m_2}{M} \,
                   \frac{1 + e^2/2}{(1 - e^2)^{5/2}} \,
                   (\sigma_1 - \sigma_2)^2 + {\cal O}(c^{-5})\;.
\end{equation}
If the companion is a white dwarf, then $\sigma_2 \simeq \alpha_0 \ll 1$. There
are quite a few pulsar -- white dwarf systems that can be utilized to constrain
$\sigma_1$ via a $\dot{P}_b$ measurement. As $\sigma_1$ is highly mass
dependent, it is important to perform such a test for different pulsar masses.
The best such constraints to date are (95\% confidence, sorted according to
pulsar mass),
\begin{itemize}
\item PSR~J1141$-$6545 ($1.27\,M_\odot$):
      $|\sigma_1 - \alpha_0| < 0.004$ \cite{bbv08},
\item PSR~J1738+0333   ($1.46\,M_\odot$):
      $|\sigma_1 - \alpha_0| < 0.002$ \cite{fwe+12},
\item PSR~J1909$-$3744 ($1.54\,M_\odot$):
      $|\sigma_1 - \alpha_0| < 0.008$,
\item PSR~J1012+5307   ($\sim 1.7\,M_\odot$):
      $|\sigma_1 - \alpha_0| < 0.008$ \cite{lwj+09},
\item PSR~J0348+0342   ($2.01\,M_\odot$):
      $|\sigma_1 - \alpha_0| < 0.005$ \cite{afw+13}.
\end{itemize}
For simplicity, we have omitted the error on the pulsars' masses in the above
list. The limit for PSR~J1909$-$3744 has been derived here for the first time,
based on the timing results recently published in Ref.~\cite{dcl+16}. Although
we have used a particular class of scalar-tensor theories, to illustrate the
physics behind these tests, the limits above can be seen more generically as
limits on any difference in gravitational charges, which leads to a rotating
gravitational dipole \cite{gw02}. In Refs.~\cite{ybyb14,ybby14,ybby14e} this has
been used to constrain two particular classes of vector-tensor theories of
gravity, that violate boost invariance in the gravitational sector. More details
are given in Section~\ref{sec:lli}.

\section{Local Position Invariance}
\label{sec:lpi}

Ernst Mach's concept of inertia was an important guiding factor in Einstein's
development of GR, and it was given a name, {\it Mach's principle}~\cite{lm07}.
The statement of Mach's principle is rather vague and generally understood as
that {\it inertia originates in a kind of interaction between
bodies}~\cite{mtw73} or {\it local physical laws are determined by the
large-scale structure of the universe}~\cite{he73}. In terms of gravity
theories, GR, although influenced by Mach's principle, is not fully Machian ---
and to some extent even appears anti-Machian \cite{rin94} --- while, for
example, the Jordan-Fierz-Brans-Dicke theory~\cite{jor55,fie56,bd61} and its
generalizations~\cite{de93,de96a,de92}, with an effective gravitational constant
depending on spacetime via a scalar field, are more Machian~\cite{bp95}. Bondi
and Samuel listed eleven versions of Mach's principle, numbered as {\sc Mach0}
to {\sc Mach10}~\cite{bs97}, wherein {\sc Mach1}, {\sc Mach3}, and {\sc Mach6}
are directly relevant to the topic here.  

Although we are not necessarily saying that Mach's principle is correct,
nevertheless, it provides interesting ideas, stimulating the exploration of
gravity theories. Especially the idea that the local gravitational law is
influenced by the universal matter distribution was considered by many
authors~\cite{whi22,sci53,bd61,will73,rai75,will93,gw08}. As we know from
cosmology, the distribution of matter is uniform and isotropic only to a certain
degree of approximation. If Mach's principle holds in a strong sense, we might
expect that the slight asymmetries in the distribution at large would result in
a slight anisotropy in the gravitational law~\cite{cs58,cs60}. In other words,
the law to interpret gravitational experiments will depend on where the
experiments are performed, with the local position invariance of gravity
violated.  Such a possibility is studied below in the context of post-Newtonian
gravity.

We concentrate on one of the ten PPN parameters which characterizes a possible
anisotropy in the gravitational interaction of localized systems due to the
universal matter distribution. Such an anisotropy is described by the Whitehead
parameter, $\xi$, in the post-Newtonian limit~\cite{will93,will14}. The $n$-body
Lagrangian has a $\xi$-term through three-body interactions (see Eq.~(6.80) in
Ref.~\cite{will93}),
\begin{equation}
  \label{eq:xi:lagrangian}
  L_\xi = - \frac{\xi}{2} \, \frac{G^2}{c^2} \sum_{i} \sum_{j\neq i} \frac{m_i
  m_j}{r_{ij}^3} \, {\bf r}_{ij} \cdot \left[\sum_{k} m_k \left( \frac{{\bf
  r}_{jk}}{r_{ik}} - \frac{{\bf r}_{ik}}{r_{jk}} \right) \right] \,.
\end{equation}
In the discussions below, the major contribution of the third body comes from
our Galaxy~\cite{will93,gw08,sw13}.\footnote{For the case of Earth tides, the
  second major contribution, from the Sun, is also included. The relative
  strengths are characterized by the gravitational potentials at the location of
  the Earth, with $U_{\rm G}/c^2 \sim 5\times10^{-7}$ for the Galaxy and
  $U_\odot/c^2 \sim 1\times10^{-8}$ for the Sun~\cite{will93}.}  We consider a
  system that the Galactic center lies in its direction of ${\bf n}_{\rm G}$ and
  at a distance of $R_{\rm G}$.  We will assume that the matter in the Galaxy
  is concentrated at the Galactic center. It was calculated explicitly that
  such an assumption, compared with the extended matter distribution in reality
  (with the bulge, the disk and the dark matter halo), introduces a correction
  factor of about two~\cite{gw08,sw13}.

The phenomena of local position invariance violation, introduced by the dynamics
deduced from Eq.~(\ref{eq:xi:lagrangian}), include anomalous Earth tides,
anomalous advance rate in planetary perihelion~\cite{will73,will93}, anomalous
precession of a circular binary orbit~\cite{swk15}, and anomalous precession of
a massive body's spin~\cite{nor87,sw13}.

{\bf Anomalous Earth Tides.}  When discussing Earth tides, it is convenient to
attribute the violation in local position invariance, characterized by $\xi$ in
the PPN formalism, to an effectively anisotropic gravitational constant (see
Eq.~(6.75) in Ref.~\cite{will93}),
\begin{eqnarray}
  G_{\rm local} &=& G_0 \left[ 1 + \xi \left( {\bf e} \cdot{\bf n}_{\rm G}
    \right)^2\left(1-\frac{3I_\oplus}{M_\oplus R_\oplus^2}\right) U_{\rm G}
    \right. \nonumber\\
  && \left. + \xi \left( {\bf e} \cdot{\bf n}_\odot
  \right)^2\left(1-\frac{3I_\oplus}{M_\oplus R_\oplus^2}\right) U_\odot \right] \,,
  \label{eq:xi:G}
\end{eqnarray}
where $I_\oplus$, $M_\oplus$, and $R_\oplus$ are the moment of inertia, mass and
radius of the Earth, respectively, $G_0$ is the bare gravitational
constant,\footnote{We have renormalized $G_0$ with an isotropic term,
  $\xi(3+I/MR^2)(U_{\rm G}+U_\odot)$, that is irrelevant to the discussions 
  here.}
  and ${\bf e}$ is the unit vector pointing from the center of the Earth to the
  location where $G$ is being measured~\cite{will93}.  The changing of ${\bf
  e}$, due to the Earth's rotation, and the changing of ${\bf n}_\odot$, due to
  the Earth's revolving around the Sun, make the gravitational interaction
  change at specific frequencies with characteristic phases~\cite{will73}.

Will~\cite{will93} summarized the primary effects of Eq.~(\ref{eq:xi:G}) on the
local gravitational acceleration: i) semi-diurnal variations with periods around
12 hours, ii) diurnal variations with periods around 24 hours, iii) long-period
zonal variations with periods of one-half year or one year, iv) long-period
spherical variations with a period of one year that, in contrast to the other
three, have no dependence on the latitude. These phenomena were called {\it
Earth's tides}, and could be studied with gravimeter~\cite{nw72,will73,will93}.
Warburton and Goodkind used data from superconducting gravimeters and obtained a
limit of $|\xi| \lesssim 10^{-3}$~\cite{wg76}. New developments in
superconducting gravimeter networks are expected to probe $\xi$ at the level of
$10^{-5}$~\cite{shi08}.

{\bf Anomalous Advance of Planetary Perihelion.} In the case of two-body
problem, Eq.~(\ref{eq:xi:lagrangian}) produces an extra acceleration between
masses $m_1$ and $m_2$ (see Eq.~(8.73) in Ref.~\cite{will93} with a different
sign convention),
\begin{equation}
  \label{eq:xi:acc}
  {\bf a}_\xi = \xi \frac{U_{\rm G}}{c^2} \frac{G(m_1+m_2)}{r^2} \left[ 2({\bf
  n}_{\rm G} \cdot {\bf n}){\bf n}_{\rm G} -3{\bf n}\left({\bf n}_{\rm G} \cdot
{\bf n}\right)^2 \right] \,,
\end{equation}
where $r$ is the coordinate separation, and ${\bf n}$ is the unit vector of
relative position.  Such an acceleration contributes to the advance rates of
planetary perihelion (see Eq.~(25) in Ref.~\cite{will73} for the expression).
Numerically, for Mercury one has an anomalous advance rate~\cite{will93},
\begin{equation}
    \delta \dot{\omega}_{\mercury} \simeq 63\, \xi \mbox{~~~arcsec
    century}^{-1}\,,
\end{equation}
while its general-relativistic advance rate is $43$\,arcsec\,century$^{-1}$.  
Because that, on one hand, other
PPN parameters also have contribution to the advance rate, and on the other
hand, the measurement precision of $\dot\omega$ is limited, $\xi$ can not be
constrained tightly from the anomalous advance of planetary
perihelion~\cite{will93}.

{\bf Anomalous Orbital Precession of Binary Pulsars.} It was shown that for
circular orbits, Eq.~(\ref{eq:xi:acc}) would lead to a precession of the orbital
norm, around the direction of the binary's acceleration towards the Galactic
center, ${\bf n}_{\rm G}$~\cite{swk15}. In pulsar timing, such a precession
causes a change in our viewing of the binary orbit, that results in a nonzero
drift in the projected semi-major axis of the pulsar orbit, 
$x \equiv a_p \sin i / c$, due to a varying
inclination angle $i$. Therefore, with the fitting parameter $\dot x$ from times
of arrival of pulse signals, the $\xi$ parameter can be constrained.  However,
the full geometric orientation of the binary orbit is required in the
calculation, while unfortunately the longitude of ascending node, $\Omega$, is
in general not an observable in pulsar timing~\cite{lk04}. With a probabilistic
assumption that $\Omega$ is uniformly picked from $[0^\circ, 360^\circ)$ in the
  sense of a {\it nuisance parameter} in Bayes' priors, the following limit was
  obtained by combining observations from PSRs~J1012+5307 and J1738+0333 at 95\%
  confidence level~\cite{swk15},
\begin{equation}
|\xi| < 3.1 \times 10^{-4} \,.
\end{equation}
In calculating the above limit, an assumption that $\xi$ has only weak
dependence on the component masses of binary pulsars was made.

{\bf Anomalous Spin Precession of the Sun.}  Nordtvedt considered
Eq.~(\ref{eq:xi:lagrangian}) in the case of extended bodies. He showed that an
isolated massive star, with internal equilibrium, undergoes a free spin
precession around its Galactic acceleration towards the Galactic center, with an
angular frequency~\cite{nor87},
\begin{equation}\label{eq:xi:spin}
\Omega^{\xi} = \xi \left( \frac{2\pi}{P}\right) U_{\rm G} \cos\psi \,,
\end{equation}
where $P$ is the star's spin period, and $\psi$ is the angle between the star's
spin and ${\bf n}_{\rm G}$.

\begin{figure}[H]%
  \centering
  \includegraphics[width=9cm,angle=0]{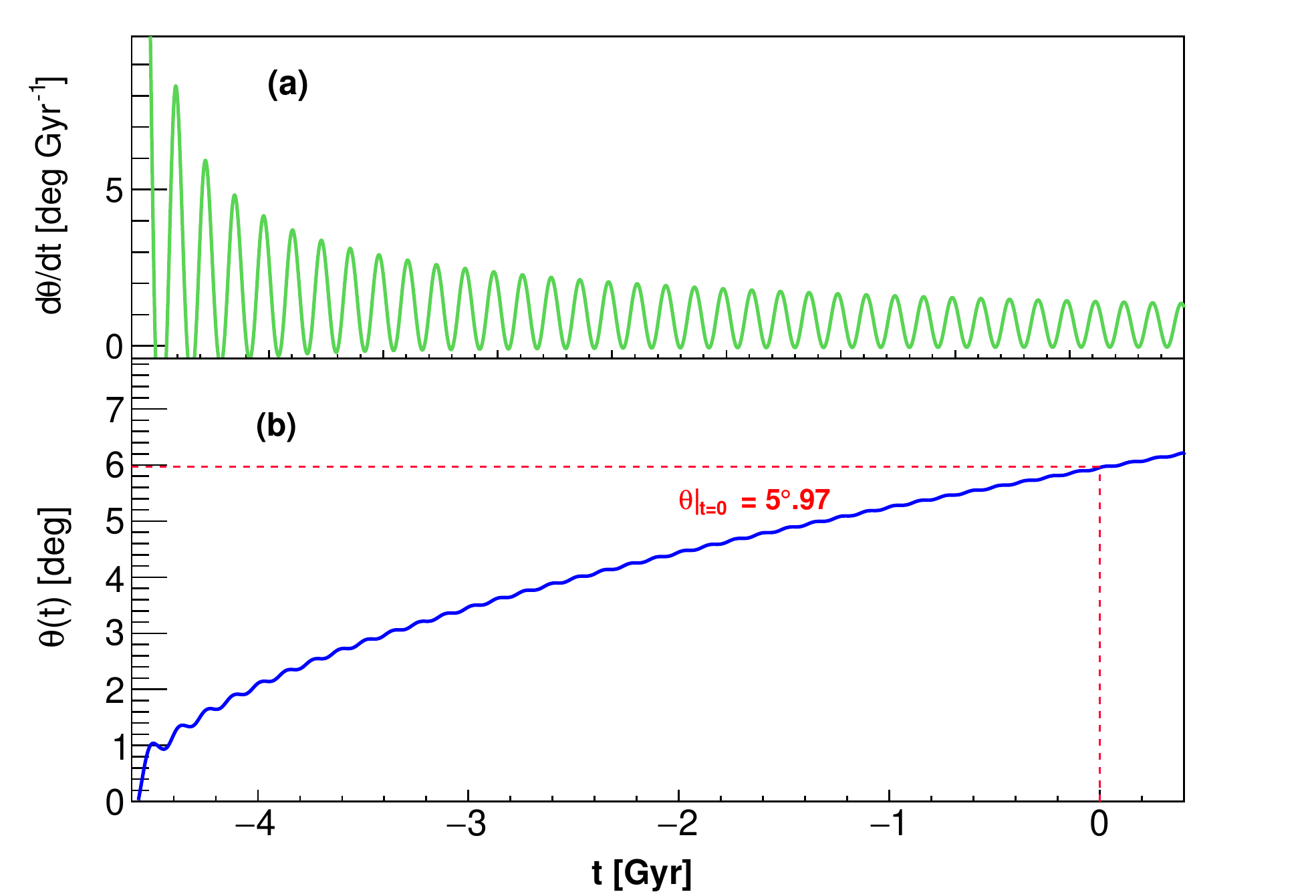}%
  \caption{A hypothetical scenario that the observed $\theta$ today ($\theta
    \simeq 5^\circ.97$) was caused purely from a nonzero $\xi$ (here $\xi= 1.12
    \times 10^{-6}$). The time derivative of $\theta(t)$, ${\rm d}\theta / {\rm
    d}t$ in (a), and $\theta(t)$ in (b) are shown as a function of time from
    the birth of the Solar System ($t\simeq -4.6$ Gyr) to today ($t=0$).
    Oscillation is caused by the Sun's revolving around the Galactic center
    ($\sim 20$ circles in 4.6 Gyr), while the decreasing amplitude in (a) is due
    to the decreasing rate of Solar spin~\cite{sku72}. } %
\label{fig:xi:solar}%
\end{figure}

From our understanding of planetary systems, at the birth of the Solar System
$\sim4.6$ Gyr ago, the angle $\theta$ between the Sun's spin and the total
angular momentum of the Solar System was likely very small. Afterwards, the
Newtonian torque is negligibly weak to change $\theta$ significantly. Today's
observations reveal a small $\theta \simeq 5.97^\circ$~\cite{aab+09}. Nordtvedt
suggested to use the observed $\theta$ as an upper limit for $\xi$-induced
precession angle during $\sim4.6$\,Gyr to constrain $\xi$ in
Eq.~(\ref{eq:xi:spin}). Detailed studies, taking the Sun's revolving around the
Galactic center into account, were carried out recently~\cite{sw13}.
Figure~\ref{fig:xi:solar} illustrates a hypothetical scenario where a nonzero
$\xi$ increases $\theta$ gradually from $0^\circ$ to its current value.  In
contrast to a constant Solar spin rate that was used in Ref.~\cite{sw13}, we
here adopt the suggestion in Ref.~\cite{ior14} and use the Skumanich law, $P
\propto (t-t_0)^{-1/2}$, for the evolution of spin rate~\cite{sku72}. A simple
calculation shows that, the adoption of Solar spindown model tightens the limit
of $|\xi|$ in Ref.~\cite{sw13} by a factor of two.  

More simulations were performed in Ref.~\cite{sw13}. It showed that, by a rough
conjecture that $\theta$ was less than $10^\circ$ at the birth of the Solar
system, $\xi$ can be limited to~\cite{sw13},
\begin{equation}
    |\xi| < 5 \times 10^{-6} \,.
\end{equation}
This value was the one obtained in Ref.~\cite{sw13} with a constant Solar spin
rate. As shown, it improves by a factor of two if the Solar spindown is
included.

{\bf Anomalous Spin Precession of Solitary Pulsars.} For solitary pulsars,
analogous to the Sun, local position invariance violation also produces a free
precession of the pulsar spin around its Galactic acceleration with the angular
frequency in Eq.~(\ref{eq:xi:spin}). For millisecond pulsars the effect is
greatly enhanced, compared to the Sun, due to the extremely short rotational
period $P$ ($\sim 10^{-3}$\,s). The rotation period of the Sun is in contrast
about one month ($\sim 10^6$\,s). Therefore, even with observations over just
ten years (compared to the $\sim4.6$\,Gyr time-scale in case of the Sun), one
can tightly constrain $\xi$ \cite{sw13}.

\begin{figure}[H]
  \centering \includegraphics[width=9cm]{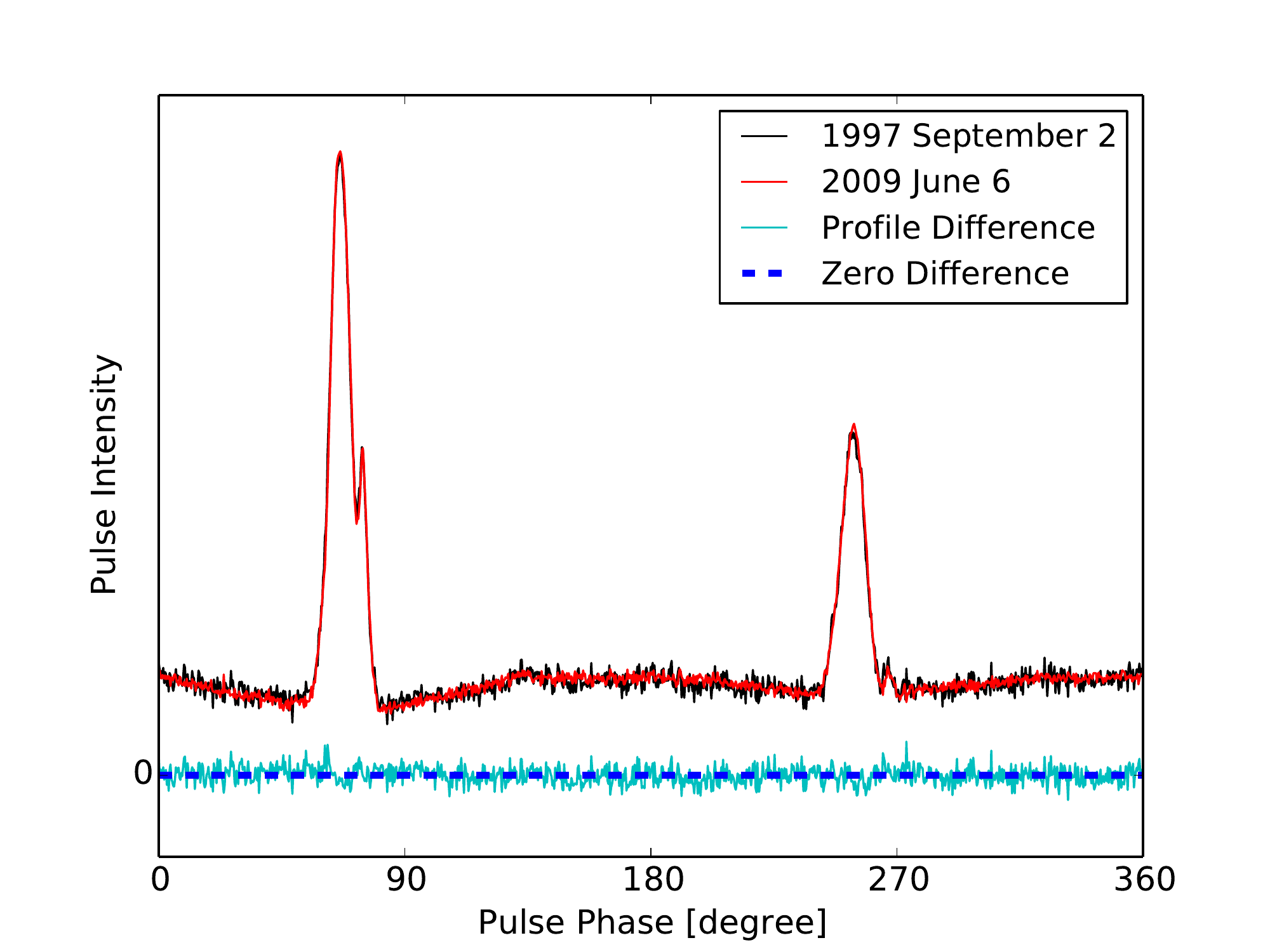}
  \caption{\label{fig:B1937}
  Comparison of two pulse profiles of PSR~B1937+21 obtained with the
  Effelsberg-Berkeley Pulsar Processor at two different epochs --- the black one
  was obtained on 2 September 1997, while the red one was obtained on 6 June
  2009. The difference between two profiles is shown in cyan at the bottom.
  Data of profiles are taken from Ref.~\cite{sck+13}.} 
\end{figure}

For millisecond pulsars, a hypothetical precession of its spin would cause
changes in our line of sight's cutting on its radiation beam. Consequently, we
would observe changes in pulse profiles as a function of time. Shao et al.\ used
data from the 100-m Effelsberg radio telescope to study the
stability of pulse profiles~\cite{sck+13}. A coherent dedispersion backend,
the Effelsberg-Berkeley Pulsar Processor, was used, which is the longest-running
coherent dedispersion backend dedicated to high-precision pulsar timing in the
world, making the database uniquely suited for the task. Hundreds of pulse profiles
of PSRs~B1937+21 and J1744$-$1134, spanning more than ten years, were analyzed
homogeneously. Detailed studies showed no pulse profile variations in two
pulsars to a high degree~\cite{sck+13}. For example, the profile of PSR B1937+21
showed variation in its width at 50\% intensity of the main pulse to be less
than a few thousandth degrees per year (see Figure~\ref{fig:B1937} for
an illustration).

\begin{figure}[H]
  \centering
  \includegraphics[width=9cm]{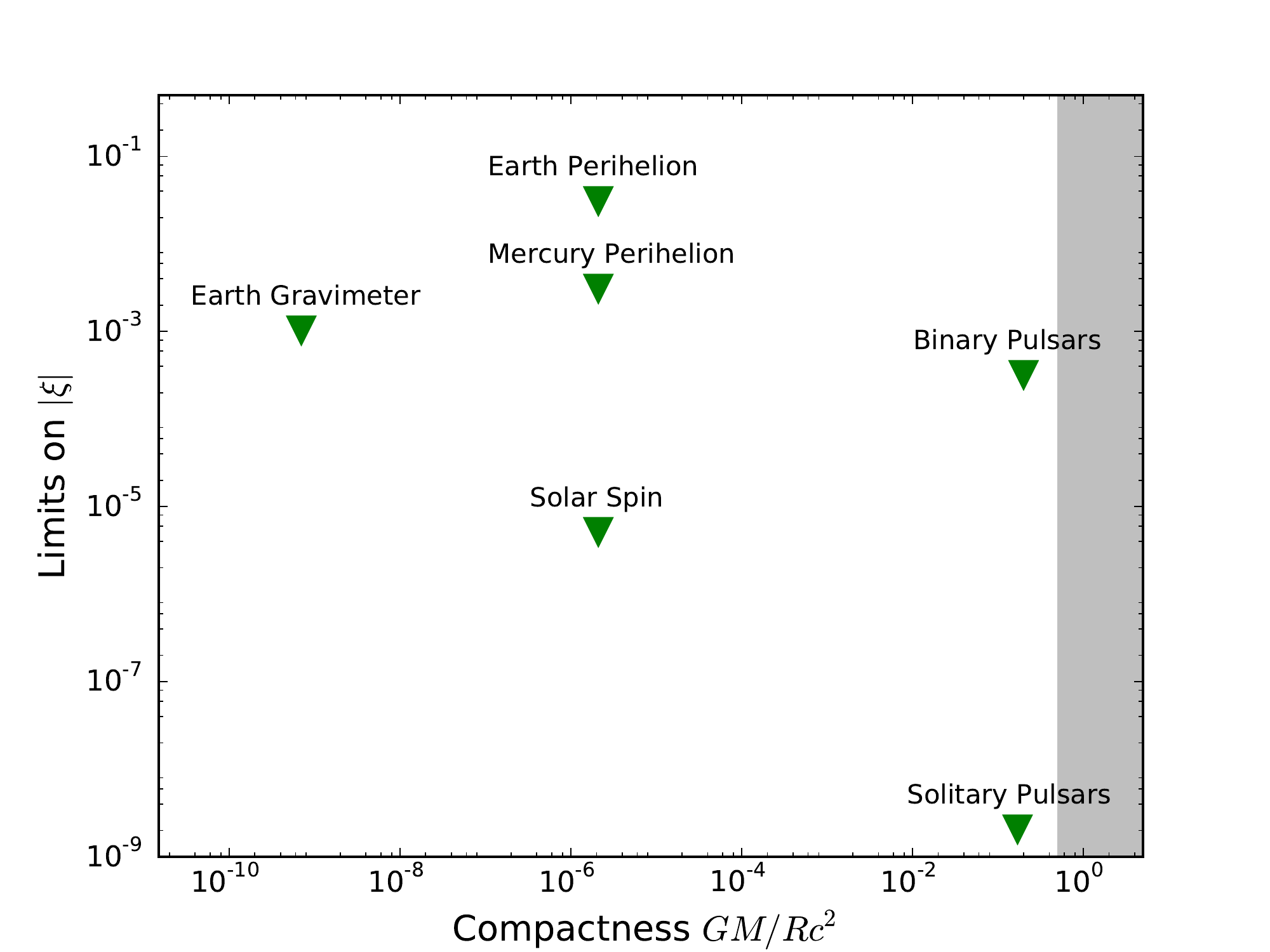}
  \caption{ Upper limits on $|\xi|$ from Earth gravimeter~\cite{wg76}, Mercury
  and Earth perihelions~\cite{will93}, Solar spin~\cite{nor87,sw13}, binary
  pulsars~\cite{swk15}, and solitary pulsars~\cite{sw13}, as a function of
  compactness, $\varepsilon \sim GM/Rc^2$, where $M$ and $R$ are mass and
radius, respectively, of the heavier body in the system. Shaded region is the
interior of black holes.}
  \label{fig:xi:limit} 
\end{figure}

The {\it null detection} of profile variation allowed to limit the precession in
Eq.~(\ref{eq:xi:spin}) for millisecond pulsars.  However, the 3-dimensional spin
orientation of the pulsar, which is needed in the calculation, is
observationally not fully constrained (although combinations of radio and
$\gamma$-ray observations can provide partial information, see e.g.,
Ref.~\cite{gjv+12}).  Therefore, a probabilistic assumption, similar to the case
of binary pulsars, was made to account for an unknown angle. With a cone model
approximating the radiation pattern~\cite{ggr84,lk04}, PSRs~B1937+21 and
J1744$-$1134 give at 95\% confidence levels~\cite{sw13}
\begin{eqnarray}
|\xi| &<& 2.2 \times 10^{-8} \,, \\
|\xi| &<& 1.2 \times 10^{-7} \,,
\end{eqnarray}
respectively. The probability distribution functions of $\xi$ (see Figure 4 in
Ref.~\cite{sw13}) have long tails that result from the probabilistic assumption
about the unknown angle. By combining the probability distribution functions of
two pulsars, a much tighter limit was obtained~\cite{sw13},
\begin{eqnarray}
|\xi| < 3.9 \times 10^{-9} \,,
\end{eqnarray}
at 95\% confidence level. This is currently the best limit on local position
invariance violation in the gravitational interaction~\cite{will14}. It can also
be converted to an upper limit on the spatial anisotropy of the gravitational
constant through Eq.~(\ref{eq:xi:G}),
\begin{equation}
\left|\frac{\Delta G}{G}\right|^{\rm anisotropy} < 4 \times
10^{-16} \quad \mbox{(95\% CL)} \,. 
\end{equation}

Figure~\ref{fig:xi:limit} summarizes the limits on $\xi$ discussed above, as a
function of the compactness of the system, $\varepsilon \sim GM/Rc^2$, where $M$
and $R$ are the mass and the radius of the system.  In the case of binary
systems, the quantities of the heavier body are used.  For pulsars with unknown
mass and/or radius, canonical values of $M \sim 1.4 \,{\rm M}_\odot$ and $R \sim
12\,{\rm km}$ are assumed.  The figure clearly shows that the limit from
solitary pulsars~\cite{sw13} not only poses the tightest constraint, but also
probes certain strong-field aspects that are not available in experiments with
weakly self-gravitating bodies.

\section{Local Lorentz Invariance}
\label{sec:lli}

The possibility to break local Lorentz invariance found great interest in the
gravity community
recently~\cite{jm01,kos04,bk06,bps11,hor09,will14,bl14,ybby14,ybby14e,tas14}.
Theoretically, it is mainly driven by the attempts to quantize the gravitational
interaction. Some models from string theory and loop quantum gravity predict a
possible breakdown of the Lorentz symmetry with novel dynamics of
spacetime~\cite{ks89a,ks89b,gp99}.  As mentioned before, Ho{\v r}ava-Lifshitz
gravity tries to construct a power-counting renormalizable, UV complete gravity
theory, at the cost of Lorentz symmetry breaking by using different scalings to
the temporal and spatial decomposition of spacetime~\cite{hor09,bps11}.  For
these theories, even at low energy scales, the full Lorentz symmetry is not
restored.  Lorentz symmetry breaking  will manifest tiny deviations in our
precision experiments, thus it probably provides us with a precious {\it quantum
gravity window}~\cite{ame13,bl14}.  In addition to the above mentioned
theoretical suspicion, experimentally exploring the boundary of fundamentally
treasured symmetries is always considered, and historically verified, as one of
the best tools in searching for new physics beyond our current
understanding~\cite{will14}.  

The breakdown of Lorentz invariance  can already happen in flat spacetime where
the Lorentz symmetry of special relativity is altered. This in turn breaks
Einstein's equivalence principle (EEP).  However, under the circumstances where
EEP is respected (hence the Lorentz symmetry of special relativity is fully
preserved), the local Lorentz invariance in gravitational interaction can still
be violated~\cite{will93}.  We are interested here in the latter possibility.
Experimental phenomena of local Lorentz invariance violation in gravity were
explored vastly at different length scales, from laboratory short-range gravity
at $\sim\mu$m~\cite{lk15} to the cosmological scale~\cite{sb14}. This review is
not intended to cover the whole field, but rather to highlight the contributions
from pulsar astronomy to tests of local Lorentz invariance at length scales of
the astronomical unit (AU; 1\,AU~$\simeq 1.5 \times 10^{11}$\,m), and with
possible strong-gravity effects associated with neutron
stars~\cite{de92b,sta03,sfl+05,lor08,gsf+11,sw12,wex14,shao14,man15}.  Two
popular frameworks with local Lorentz invariance violation in gravity are
considered in the following.

\subsection{Metric-based framework: PPN}
\label{sec:lli:ppn}

In the field of experimental gravity, the most popular framework is the
parametrized post-Newtonian (PPN) formalism, proposed by Kenneth Nordtvedt and
Clifford Will in late 1960s and early 1970s~\cite{wn72,nw72}. The construction
of PPN formalism was inspired by the experimental precision of WEP at that time
and thence the requirement of a metric theory of gravitation that fulfills the
EEP~\cite{will14}. In metric theories of
gravitation, despite the possible existence of long-range gravitational fields
(e.g., scalar fields in scalar-tensor theories~\cite{fm03}), matter and all
non-gravitational fields only couple to the (physical) metric
$g_{\mu\nu}$~\cite{will93,will14}. For instance, the motion of test particles
follows the geodesics of the geometry, described by the symmetric rank-2 metric
tensor. Consequently, all test bodies, independent of their composition, fall
in the same way, which means that WEP is fulfilled. More generally, metric theories
of gravity fulfill the EEP, which is also one of the building blocks for PPN
formalism.

In the PPN framework, generic ways of how the metric depends on the
energy-momentum content of matters are considered. In the PPN gauge, the
dependence of the metric components on the matter content can be found in the
classical reviews by Will~\cite{will93,will14}.  Ten PPN parameters control, at
the first post-Newtonian order, the couplings between the metric and various
matter potentials, including the well-known Newtonian ``scalar'' potential, 
and vectorial/tensorial ones.  Ten PPN parameters are
intelligently organized to describe different aspects of gravitation.  For
example, the Eddington-Robertson-Schiff parameters, $\beta$ and $\gamma$,
describe the nonlinearity in the superposition law of gravity and the amount of
space-curvature produced by a unit rest mass, respectively. PPN parameters,
$\alpha_1$, $\alpha_2$ and $\alpha_3$ parametrize different effects related to a 
gravitationally preferred frame, which is to be discussed here ($\alpha_3$ is in 
addition
related to energy-momentum conservation; see below). From the last forty years,
there are quite some remarkable limits from various experiments. In this
subsection, to keep the scope of this review controllable, we will focus on the
constraints of $\alpha_1$, $\alpha_2$, and $\alpha_3$ from radio pulsars, while
interested readers are encouraged to read Refs.~\cite{will93,will14} for more
details.  

{\bf Orbital polarization of binary pulsars with a preferred frame.} We will
discuss the ``orbital polarization'' phenomenon introduced by a nonzero
$\alpha_1$, and the relevant experimental tests from binary pulsars.  In the
generic framework of PPN formalism, there could be a preferred frame where the
gravitational interaction is isotropic. If a system is moving with respect to
this rest frame, its ``absolute'' movement is manifest in the gravitational
interactions inside the frame that is attached to the system.  Equivalently
speaking, the gravitational interaction of an isolated system of masses depends 
on the movement of the frame.  Such a preferred frame could have been formed at 
a very early stage in the cosmological evolution of the
Universe or from local matter distributions. The widely adopted practice is to
use the frame where the cosmic microwave background (CMB) is isotropic at first
order approximation. We will call it the CMB frame.

We consider the orbital dynamics of a two-body system, using a binary pulsar as
a prototype.  If a binary pulsar is moving with respect to such a preferred
frame with an absolute velocity, {\bf w}, its orbital dynamics is determined by
the post-Newtonian Lagrangian, $L$, where, in addition to the contributions from
Newtonian gravity and general-relativistic corrections, we have extra
contributions from $\alpha_1$ and $\alpha_2$\footnote{We ignore the
  contributions from $\beta$ and $\gamma$ for the moment, and postpone the
contributions from other PPN parameters.},
\begin{eqnarray}
    L_{\alpha_1} &=& -\alpha_1 \frac{Gm_1m_2}{r} \frac{{\bf v}_1^0 \cdot {\bf
    v}_2^0}{2c^2} \,, \label{eq:ppn:a1} \\
    L_{\alpha_2} &=& \alpha_2 \frac{Gm_1m_2}{r}
    \frac{\left({\bf v}_1^0 \cdot {\bf v}_2^0 \right)-\left({\bf n} \cdot {\bf
    v}_1^0\right) \left({\bf n} \cdot {\bf v}_2^0 \right)}{2c^2} \,,
    \label{eq:ppn:a2}
\end{eqnarray}
where ${\bf v}_i^0$ ($i=1,2$) is the absolute velocity of body $i$.

\begin{figure}[H]
    \centering
    \includegraphics[width=8cm]{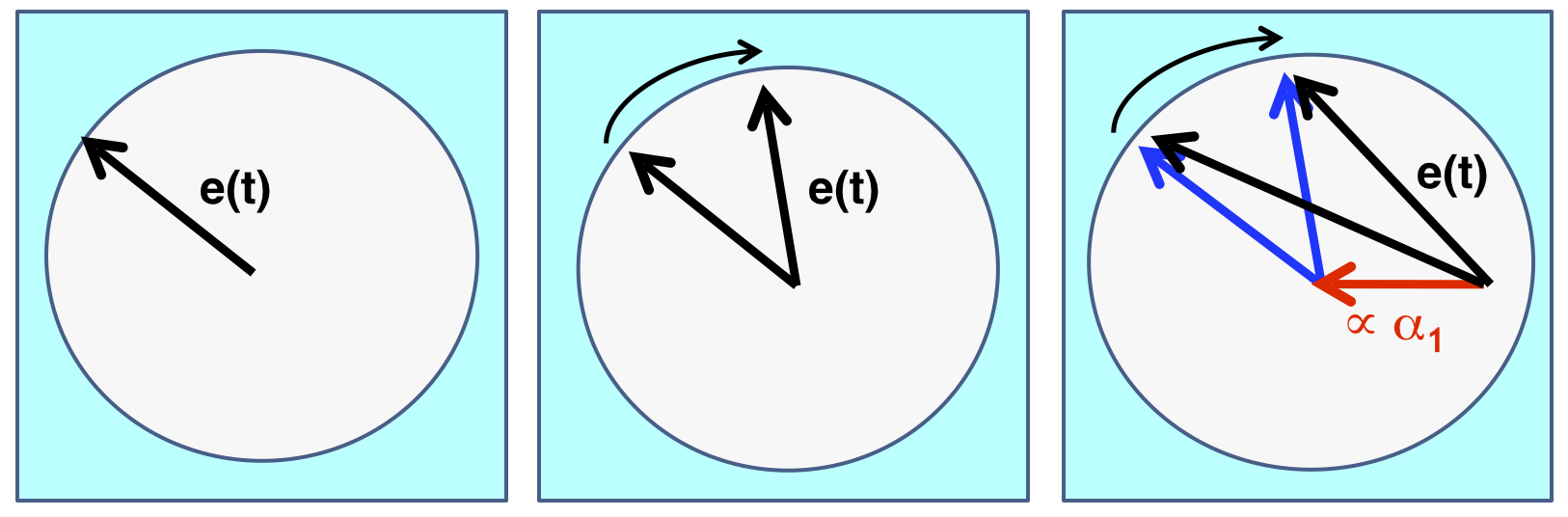}
    \caption{\label{fig:ecc:a1} Graphical illustration of the evolution of the
    eccentricity vector, ${\bf e}(t)$, for quasi-circular orbits in ({\it left})
    Newtonian gravity, ({\it middle}) general relativity, and ({\it right}) PPN
    formalism with a nonzero $\alpha_1$. Black arrows, ${\bf e}(t)$, represent
  the eccentricity vectors in observation.}
\end{figure}

Damour and Esposito-Far{\`e}se are the first to study the possibility of local
Lorentz invariance violation in PPN formalism within the context of pulsar
timing~\cite{de92b}. Because of the tight limit of $\alpha_2$ from spin
evolution of the Sun (see below) at that time~\cite{nor87}, they discarded the
term proportional to $\alpha_2$, namely Eq.~(\ref{eq:ppn:a2}). With the
contribution of Eq.~(\ref{eq:ppn:a1}), Damour and Esposito-Far{\`e}se advised a
pictorial way to capture the secular influence of $\alpha_1$ on the dynamics of
the quasi-circular orbit. The effect is included in the evolution of the orbital
eccentricity vector, ${\bf e}\equiv e{\bf a}$, where $e$ is the magnitude of the
orbital eccentricity and ${\bf a}$ is the unit vector pointing from the center
of mass of the system to the orbital periastron.  The evolution of ${\bf e}$ is
illustrated in Figure~\ref{fig:ecc:a1} and explained as follows in different
gravity theories.

\begin{itemize}
    \item In Newtonian gravity, it is well known that the $\propto r^{-2}$ force
      has the symmetry group of $O(4)$ instead of $O(3)$, therefore, the unit
      vector ${\bf a}$ is a conserved quantity, hence ${\rm d}{\bf e}/{\rm d}t =
      0$.
    \item In GR, Einstein derived the famous periastron advance rate for the
      Mercury~\cite{ein15a}. In the first post-Newtonian approximation, the
      (orbital averaged) eccentricity vector ${\bf e}(t)$ rotates uniformly 
      as a function of time.
    \item In addition to GR, if a nonzero $\alpha_1$ exists, the eccentricity
      vector ${\bf e}(t)$ can be viewed as a vectorial superposition of a forced
      eccentricity, ${\bf e}_{\rm F}$, that is a constant vector in the orbital plane,
      and a rotating eccentricity, ${\bf e}_{\rm R}(t)$, that rotates according
      to the relativistic precession rate of periastron~\cite{de92b}, and is 
      constant in length. The observed  eccentricity vector
      ${\bf e}(t) = {\bf e}_{\rm F} + {\bf e}_{\rm R}(t)$.
\end{itemize}

In early work, because of the lack of suitable systems and essential
observations to perform a direct test of the above dynamics, probabilistic
approaches were adopted, analogue to the Damour-Sch\"afer test described in 
Section~\ref{sec:uff}~\cite{de92b,bcd96}.  Later the method was extended to 
statistically combine multiple systems by taking care of a potential selection 
effect when simply picking a system with the most favorable parameter 
combination~\cite{wex00}. In addition, because the 3-dimensional velocity of the 
binary pulsar systems cannot be fully obtained from radio timing, one has to 
invoke probabilistic assumptions for the radial velocity of most of the binary 
systems used in the analyses. Originally, PSRs 0655+64 and 1855+09 were used to 
obtain the first limit of $\alpha_1$ in pulsar timing experiment~\cite{de92b}.  
Later, PSR J2317+1439 was used to update its value~\cite{bcd96}.  
A direct method to measure the influence of $\alpha_1$ on orbital dynamics was
developed in Ref.~\cite{wk07}, where a timing formula including
Lorentz-violating effects from $\alpha_1$ and $\alpha_2$ is constructed for the
double pulsar system. However, with 3-year timing data of the Double
Pulsar~\cite{ksm+06}, the constraints on $\alpha_1$ and $\alpha_2$ are quite
weak. Because the orbit of double pulsar is nearly edge-on with respect to its
line of sight~\cite{ksm+06}, the timing formula gets simplified. For a generic
orbital orientation, it can be rather complicated. 

The best constraint on $\alpha_1$ comes from a new method, based on the
observations of small-eccentricity neutron star -- white dwarf
binaries~\cite{sw12}. It is a direct method that looks for time variation of the
eccentricity vector, instead of using the magnitude of the eccentricity as in
the aforementioned probabilistic approaches. The problem related to the
measurement of 3-dimensional spatial velocity is resolved by using extra
information from optical observations of the white dwarf companion. From
combined information of radio timing of the pulsar and  high-resolution
spectroscopy of the white dwarf, the 3-dimensional spatial velocity of the
binary can be obtained, therefore, the absolute velocity of the system with
respect to the CMB frame is known.  In the direct approach, all quantities that
enter the differential equations of orbital dynamics are known, except the
longitude of ascending node, $\Omega$. We performed extensive simulations to
evolve the orbit according to the post-Newtonian dynamics with a nonzero
$\alpha_1$ for all possible values of $\Omega \in [0^\circ, 360^\circ)$, and
  from simulations to infer what kind of eccentricity variation, in terms of
  both magnitude and direction, would be consistent with real observations. The
  $\alpha_1$ values that result in eccentricity variations compatible with real
  observational data are kept. From the distribution of $\alpha_1$ for every
  $\Omega$, we were able to obtain an upper limit of $\alpha_1$ as a function of
  $\Omega$. From these results, we picked the most conservative one as our final
  limit, thus thoroughly avoided the probabilistic assumption of Ref.~\cite{sw12}.

In the approach outlined above, we obtained upper limits from PSRs J1012+5307
and J1738+0333.  Both systems have multiwavelength observations. In particular
the optical observations of their white dwarf companions are very useful in
giving 3 dimensional velocity and component masses in a way that is independent
to the underlying gravity theory~\cite{wex14}.  Among other factors that affect
the test, the orbit of PSR J1738+0333 has a better relative geometrical
orientation with respect to the absolute velocity of the system in the CMB
frame. Therefore, a better limit was achieved in that system. At 95\% confidence
level, the analysis of PSR J1738+0333 gives,
\begin{equation}\label{eq:limit:J1738}
    \alpha_1 = -0.4^{+3.7}_{-3.1} \times 10^{-5} \,.
\end{equation}
It is the currently best limit of the PPN parameter $\alpha_1$~\cite{will14}.
If a similar mechanism to the cosmological attractor in scalar-tensor
theories~\cite{dn93} exists for the PPN parameter $\alpha_1$ (as for the
Eddington-Robertson-Schiff parameters $\beta$ and $\gamma$), the new limit at
the level of ${\cal O}(10^{-5})$ in Eq.~(\ref{eq:limit:J1738}) has started to
enter cosmologically interesting parameter space.

{\bf Orbital precession of binary pulsars.} As stated above, Damour and
Esposito-Far{\`e}se initiated the program of using pulsar timing to probe the
local Lorentz invariance in the gravity sector~\cite{de92b}. We followed their
work and derived the joint influence of $\alpha_1$ and $\alpha_2$ on the orbital
dynamics of an eccentric binary~\cite{sw12}. It is interesting to discover that,
for small-eccentricity binaries with $e \ll 1$, the effects of $\alpha_1$ and
$\alpha_2$ on orbital dynamics decouple. While the $\alpha_1$ parameter tries to
polarize the orbit, its effect is inside the orbit; in contrast, the $\alpha_2$
parameter has an effect ``perpendicular'' to the orbit in the sense that it
tries to rotate the orbit along the direction of the absolute velocity of the
binary, ${\bf w}$. The precession rate is~\cite{sw12},
\begin{equation}\label{eq:a2:orbit}
    \Omega_{\alpha_2}^{\rm orbit} = - \alpha_2 \frac{\pi}{P_b} \left(\frac{w}{c}
    \right)^2
    \cos\psi_{\rm orbit} \,, 
\end{equation}
where $P_b$ is the orbital period and $\psi_{\rm orbit}$ is the angle between
orbital angular momentum and the direction of ${\bf w}$.  In pulsar timing, such
a precession will change the angle between our line of sight and the orbital
plane, hence change the projected semimajor axis of the pulsar orbit, which is
an observable in timing experiments. The upper limits of the change in the
projected semimajor axis can be used to constrain such an $\alpha_2$-induced
precession.  After carefully subtracting other potential contributions to the
change of projected semimajor axis, we were able to reach a limit of,
\begin{equation}
    |\alpha_2| < 1.8 \times 10^{-4} \,,
\end{equation}
from the combination of PSRs J1012+5307 and J1738+0333 at 95\% confidence
level~\cite{sw12}. In obtaining the above quoted limit, a probabilistic
assumption about $\Omega$ has to be made, and we also made an assumption of a
weak compactness-dependence of $\alpha_2$ for the two pulsars used in the test.

{\bf Spin precession of solitary pulsars.} The limit of $\alpha_2$ obtained from
binary pulsars is much weaker than the limit from the spin evolution of the Sun,
that is at the level of ${\cal O}(10^{-7})$~\cite{nor87}. As for the $\xi$
parameter discussed above, Nordtvedt calculated the effect of $\alpha_2$ on a
rotating massive body with internal equilibrium, and found that the spin
direction of such a body would precess around the direction of its absolute
movement at a rate,
\begin{equation}\label{eq:a2:spin}
    \Omega_{\alpha_2}^{\rm spin} = - \alpha_2 \frac{\pi}{P}
    \left(\frac{w}{c}\right)^2
    \cos\psi_{\rm spin} \,,
\end{equation}
where $P$ is the rotation period of the body and $\psi_{\rm spin}$ is the angle
between spin direction and ${\bf w}$. This precession formula is very similar to
the orbital precession rate in Eq.~(\ref{eq:a2:orbit}) with replacements of $P_b
\to P$ and $\psi_{\rm orbit} \to \psi_{\rm spin}$. It is because that for a
uniformly rotating extended body, its constitution particles can be viewed as
pairs in orbit with period $P$, and the orbits of these pairs all precess with
an angular velocity $\Omega_{\alpha_2}^{\rm orbit}$, therefore the extended body
as a whole precesses with $\Omega_{\alpha_2}^{\rm spin}$.  The disproportional
gravitational forces in above analogy are compensated by the forces from the
internal pressure. 

Nordtvedt assumed that the spin of the Sun was in parallel with the total
angular momentum of the Solar system when the Solar system was formed about five
billion years ago, and used the currently observed misalignment angle
$\sim6^\circ$ as an upper limit to constrain the possible precession of the
Solar spin introduced by a nonzero $\alpha_2$~\cite{nor87}. This approach is
very effective due to a long time baseline of precession, about five billion
years, therefore, a very tight limit of $\alpha_2$ at the level of ${\cal
O}(10^{-7})$ was achieved.  

Nordtvedt pointed out briefly that observations of pulsars can also be used to
constrain $\alpha_2$~\cite{nor87} if possible observational quantities could be
firmly identified. PSRs~B1937+21 (the first discovered millisecond pulsar) and
B0531+21 (the Crab pulsar) were used as potential examples in his discussion.
Following this suggestion, Shao et al.~\cite{sck+13} were the first to firmly
connect pulsar observations to the theoretical prediction of the spin precession
introduced by a nonzero $\alpha_2$.  If such a precession exists for solitary
pulsars, it will introduce changes in our line of sight cutting the emission
region of pulsars' magnetosphere.  Therefore, we are supposed to see drifts in
pulse profiles as a function of time.  As mentioned before, pulse-profile
observations of two millisecond pulsars, PSRs B1937+21 and J1744$-$1134,
conducted at the Effelsberg radio telescope, were analyzed homogeneously to
obtain information about the possibility of such a drift.  From data collected
with the Effelsberg-Berkeley Pulsar Processor (EBPP) backend, covering more than
ten years, no detectable change in profiles is identified. Quantitative limits
on the change of widths, change of peak separations, and change of peak
intensity ratios, of pulse profiles, were obtained (see Figure~4 and Table~1 in
Ref.~\cite{sck+13}).  With a reasonable geometrical model for the pulsar
emission, the observed (non-)change in pulse profile is linked to the
(non-)precession of the pulsar spin. Analogous to the limit on $\xi$ described
in Section~\ref{sec:lpi}, new limits for $\alpha_2$ were inferred from the
absence of precession given by Eq.~(\ref{eq:a2:spin}).

In performing the test, we need the full geometrical information about the spin
direction of pulsars, and the relative movements of pulsars with respect to a
preferred frame, which, unfortunately, are not fully attainable.  Instead of the
3 dimensional velocity, we can only obtain its 2 dimensional projection on the
sky plane, namely the proper motion. However, with the knowledge of the Galactic
potential of the Milky Way, we can get a reasonable range for the velocity along
the line of sight (the unknown radial velocity), by assuming that the pulsar is
gravitationally bound to the Galaxy. Our limit of $\alpha_2$ is not very
sensitive to the radial velocity in that range.  For two angles that describe
the spin direction, we can infer one of them, the polar angle, with the help of
$\gamma$-ray observations from the Fermi satellite. The other angle, the
azimuthal one, is assumed to be uniform in the range of $[0^\circ, 360^\circ)$
  probabilistically. After taking all experimental uncertainties and the
  probabilistic assumption into account, we obtained at 95\% confidence level,
    \begin{equation}\label{eq:a2:limit}
        |\alpha_2| < 1.6 \times 10^{-9} \,,
    \end{equation}
    from the combination of PSRs B1937+21 and J1744$-$1134~\cite{sck+13}. 
    
\begin{figure}[H]
  \centering
  \includegraphics[width=9cm]{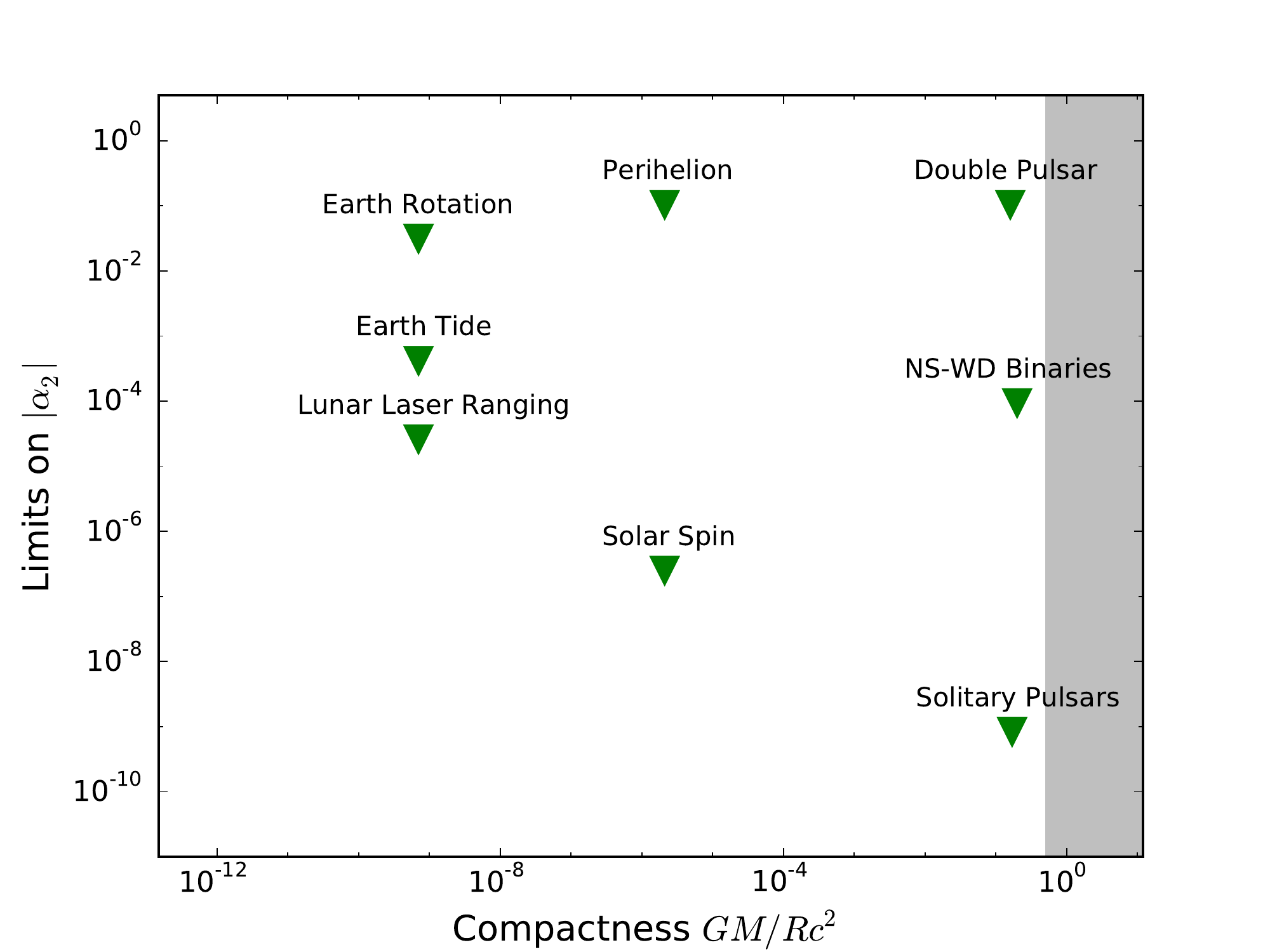}
  \caption{\label{fig:a2:limit} Upper limits on $|\alpha_2|$ from Earth
  rotation~\cite{will93}, Earth tides~\cite{wg76,will93}, planetary
  perihelion~\cite{will93}, LLR~\cite{mwt08}, Double
  Pulsar (based on data up to 2006)~\cite{wk07}, Solar spin~\cite{nor87}, 
  neutron star -- white dwarf
  binaries~\cite{sw12}, and solitary pulsars~\cite{sw13}, as a function of
  compactness, $GM/Rc^2$, where $M$ and $R$ are mass and radius, respectively,
  of the heavier body in the system. Shaded region is the interior of
  black holes.}
\end{figure}

In Figure~\ref{fig:a2:limit} we plot the limits of $\alpha_2$ from various
observations as a function of the compactness of the
system~\cite{will93,will14}.  The limit from pulse-profile observations of
pulsars not only constitutes the currently best limit of
$\alpha_2$~\cite{will14}, but also probes a parameter region that strong
gravitational fields are present.

\begin{figure}[H]
  \centering
  \includegraphics[width=9cm]{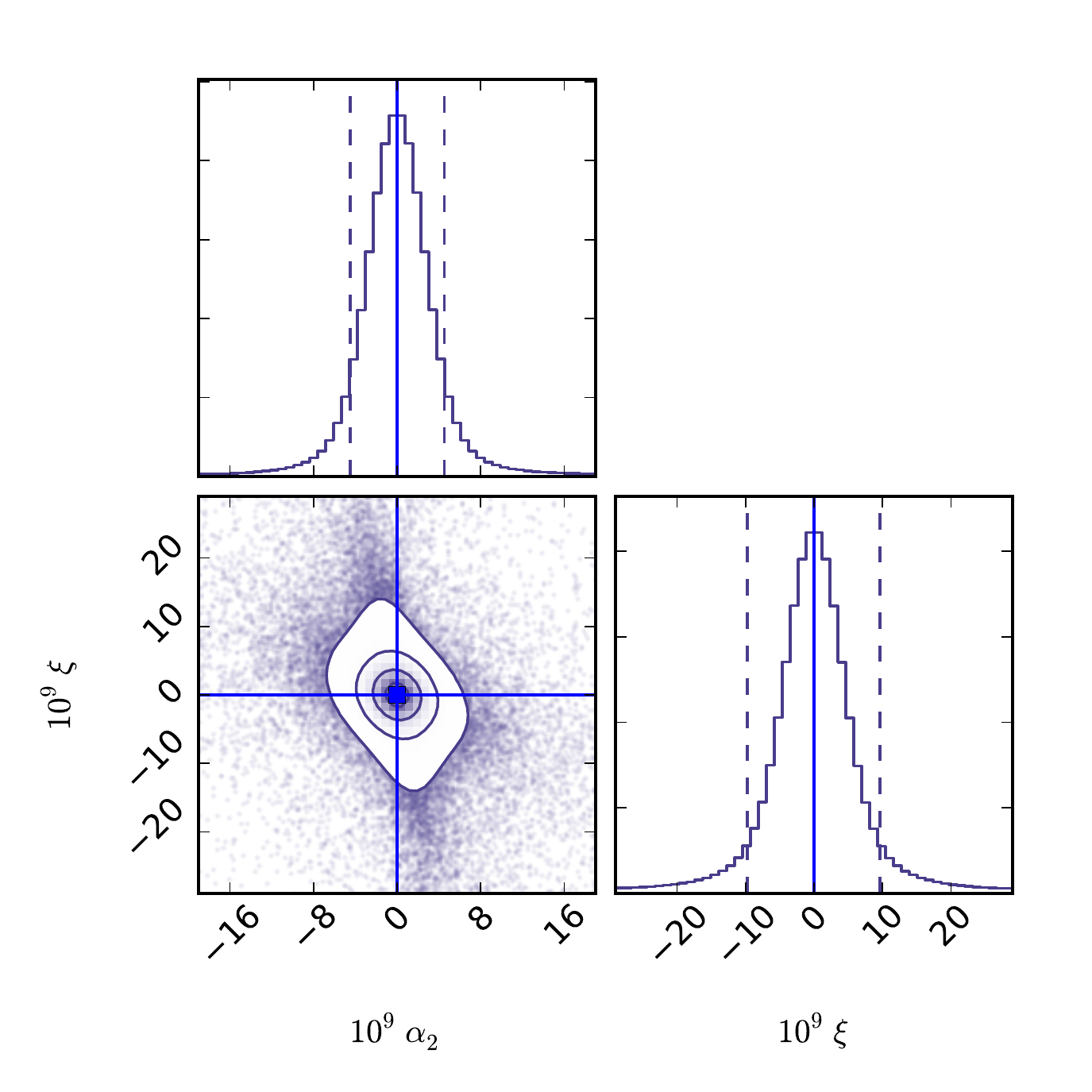}
  \caption{\label{fig:a2:xi} Distributions of $(\alpha_2,\xi)$ pairs whose
  effects on the spin precession of solitary pulsars, PSR~B1937+21 and
  J1744$-$1134, are compatible with observations. The values of $\alpha_2$ and
  $\xi$ in GR are marked in blue. The marginalized probability densities of two
parameters are shown as histograms where dash lines enclose the 90\% confidence
region.} 
\end{figure}

As discussed in detail in Section~\ref{sec:lpi}, the PPN parameter $\xi$, that
describes the local position invariance violation, introduces a similar
precession of the spin of solitary pulsars, and were constrained by the same set
of observations of PSRs~B1937+21 and J1744$-$1134 independently~\cite{sw13}.
The precessions introduced by $\alpha_2$ and $\xi$ are around different
directions (${\bf w}$ and ${\bf n}_{\rm G}$, respectively). Possible correlation
between $\alpha_2$ and $\xi$ is expected.  We performed new simulations that
include the possibility that both $\xi$ and $\alpha_2$ can be nonzero. The
result on the change in pulse widths from simulations with randomly generated
$(\alpha_2,\xi)$ pairs is compared with pulse-profile observations in
Ref.~\cite{sck+13}.  In Figure~\ref{fig:a2:xi} we plot the $(\alpha_2,\xi)$
pairs in our simulations that are compatible with the observational constraints.
The quantitative limits on the changes of pulse profiles of PSRs~B1937+21 and
J1744$-$1134 are used, and an assumption of a uniform distribution of $\eta$
in $[0^\circ, 360^\circ)$ is made. As we can see from the figure, because we
  have two observational constraints to be satisfied from two pulsars, the
  degeneracy between $\alpha_2$ and $\xi$ is broken to some degree. The
  consideration of simultaneously nonzero values for $\alpha_2$ and $\xi$ does
  not degrade too much the previous limits, based on only one nonvanishing PPN
  parameter. The change is only a factor of a few.  We also performed mock
  simulations to check further improvement of these limits by including a third
  hypothetical pulsar with the same level of quality in pulse-profile
  observation. It can improve our joint limits to the level when only one
  parameter is considered nonzero. This improvement largely comes from the
  suppression of long tails in the posterior distributions (as discussed in
  detail in Ref.~\cite{sw12}).

Again, if a similar mechanism to the cosmological attractor in scalar-tensor
theories~\cite{dn93} exists for the PPN parameter $\alpha_2$, the new limit from
pulsar pulse profiles at the level of ${\cal O}(10^{-9})$, namely
Eq.~(\ref{eq:a2:limit}), has definitely entered into a region that can be used
to constrain the attractor mechanism.  The limit is also interesting for
phenomena in cosmology with Lorentz-violating gravity~\cite{sb14}.

{\bf Orbital polarization of binary pulsars with self acceleration.} The
$\alpha_3$ parameter has double effects as it violates the local Lorentz
invariance and the energy-momentum conservation law in gravity~\cite{will14}.
The existence of a nonvanishing $\alpha_3$ could introduce interesting phenomena
with anomalous planetary perihelion precession~\cite{nw72}, anomalous time
derivative of pulsars' spin~\cite{will93}, and polarization of binary
orbits~\cite{bd96}.  In pulsar timing experiments, the best constraint of
$\alpha_3$ comes from its effects on orbital dynamics of binary
pulsars~\cite{bd96,wex00,sfl+05,gsf+11,zsd+15}.

The PPN parameter $\alpha_3$ breaks the conservation of energy-momentum,
introducing a self acceleration on spinning bodies with
self-gravity~\cite{will93,bd96},
\begin{equation}
  \label{eq:ppn:a3}
    {\bf a}_{\alpha_3} = - \frac{1}{3} \frac{E^{\rm grav}}{Mc^2} {\bf v}^0
    \times {\bf \Omega} \,,
\end{equation}
where ${\bf v}^0$ is the absolute velocity, ${\bf \Omega}$ is the spinning
angular velocity, and $E^{\rm grav}$ is the gravitational self-energy. It is
easily seen that, i) the more compact the body (hence a larger $|E^{\rm
grav}/Mc^2|$), the larger the effect of
$\alpha_3$; ii) the more rapidly spinning the body, the larger the effect of
$\alpha_3$. Millisecond pulsars represent best laboratories so far for testing
this PPN parameter.  

Both spin derivative of solitary pulsars and orbital polarization of binary
pulsars were used to constrain $\alpha_3$. Here we will emphasize the latter,
for it gave a significantly better limit.  The effect of $\alpha_3$ on the
binary orbit is very similar to the effects of GWEP violation in
Section~\ref{sec:uff} and $\alpha_1$ that we discussed before. It introduces a
``gravitational Stark effect'' that polarizes the orbital eccentricity vector
for quasi-circular neutron star -- white dwarf orbits with the pulsar spin
perpendicular to the orbital plane~\cite{bd96}.  Similar to the case shown in
Figure~\ref{fig:ecc:a1}, a nonzero $\alpha_3$ makes the observed eccentricity
vector a superposition of two vectors, like in Figure~\ref{fig:ecc:a1}, namely, a
rotating vector, precessing with the rate of the relativistic periastron
advance, and a fixed vector that is inside the orbital plane with a direction
perpendicular to both the pulsar spin and the system's absolute
velocity~\cite{bd96}. The assumption that the spin of the pulsar is aligned with
the orbital angular momentum is well justified for recycled millisecond pulsars.
In contrast to $\alpha_1$, the extra contribution with a nonzero $\alpha_3$ is
nonconservative, in the sense that it violates momentum conservation. It is a
self-acceleration effect.

With probabilistic assumptions about the relative orientation of the pulsar
orbit with respect to the absolute velocity ${\bf w}$, and the relative angle
between the rotating eccentricity and the fixed eccentricity, Bell and Damour
obtained a limit of $|\alpha_3| < 2.2\times10^{-20}$ at 90\% confidence level.
Later, it was updated several times with new
observations~\cite{wex00,sfl+05,gsf+11} and more sophisticated statistical
treatments, leading to $|\alpha_3| < 5.5 \times 10^{-20}$ at 95\% confidence
level~\cite{gsf+11}. The most recent limit comes from PSR J1713+0747, and gives
$|\alpha_3| < 2 \times 10^{-20}$ at 95\% confidence level~\cite{zsd+15}. This
remarkable limit is even a million times better than the limits on the violation
of WEP~\cite{will14}. One needs to keep in mind, however, that
this last limit is based on the Damour-Sch\"afer test applied to a single
system, and should therefore be taken with some caution, as outlined previously.
On the other hand, in the meantime, with new constraints on a temporal variation in the eccentricity, PSR J1713+0747 is very well suited for a direct test of 
$\alpha_3$ (Zhu et al., in prep.).

\subsection{Action-based framework: SME}
\label{sec:lli:sme}

Two pillars of modern theoretical physics, {\it i.e.}, the standard
model of particles and the general relativity of gravitation, can be
elegantly expressed in the language of field theory, where the action,
$S \equiv \int {\cal L} {\rm d}^n x$, with the Lagrangian density
${\cal L}$ and the spacetime dimension $n$ (we fix $n = 4$ hereafter),
plays a prominent role. Classically, the dynamics of physics is
obtained with the principle of least action, that requires a vanishing
change in the action, $\delta S = 0$, with independent variational
changes in its dynamical field variables. It results in an extremum of
the action integral and picks out the extremal paths.\footnote{Quantum
mechanically, in the language of path integral, all paths are
equal-probabilistic, but paths that are not extremal mutually cancel
out vastly because of lack of accordance in their phases, and in the
classical limit where the Planck constant $\hbar \to 0$, only the
extremum contributes.}  From a field-theoretical viewpoint, the action,
or equivalently the Lagrangian density, encapsulates the physical dynamics.

Historically, shortly after Albert Einstein established the field
equations of gravitation~\cite{ein15}, David Hilbert proposed its
corresponding action formulation~\cite{hil15} starting with the
{\it Einstein-Hilbert action}~\cite{mtw73},
\begin{equation}
  S_{\rm EH} = \frac{1}{16\pi G}\int \sqrt{-g} (R-2\Lambda)  {\rm
    d}^4 x \,, 
\end{equation}
where $R$ is the Ricci scalar and $\Lambda$ is the cosmological
constant (we set $\Lambda=0$ for localized systems). To go beyond
Einstein's GR, new gravitational degrees are added to
the above action with suitable forms, like the scalar degree in
scalar-tensor theories~\cite{de92,de93,de96a,de96b,de98} mentioned in
Section~\ref{sec:uff}.

Since GR performed extraordinarily well in a variety of experiments, only tiny
deviations from GR are allowed in the regimes that have been probed, if assuming
the absence of nonperturbative strong-field dynamics.  Therefore, effective
field theories with effective degrees that come from deeper fundamental
(unknown) theory fit the context. Standard model extension (SME) is constructed
as a convenient experimentally working framework to probe all possible
Lorentz-violating deviations in the spirit of effective field
theories~\cite{ks89a,ks89b,ck97,ck98}. In the limit of Riemannian spacetime and
the pure-gravity sector with Lorentz-violating operators of 
mass dimension four or less (the
so-called {\it minimal} SME), the only possible Lagrangian with gauge
invariance, energy-momentum conservation, and Lorentz-covariant dynamics,
is~\cite{kos04,bk06}
\begin{eqnarray}
  S &=& \frac{1}{16\pi G} \int \left[\sqrt{-g}\left(R-2\Lambda - uR +
    s^{\mu\nu}R_{\mu\nu}^{\rm T} +
    t^{\kappa\lambda\mu\nu}C_{\kappa\lambda\mu\nu} \right) \right. 
    \nonumber \\ 
    && \left. - V(\cdot)
    \right] {\rm d}^4 x \,,\label{eq:sme:lagrangian}
\end{eqnarray}
where $R_{\mu\nu}^{\rm T} \equiv R_{\mu\nu} - g_{\mu\nu}R/4$ is the
traceless Ricci tensor and $C_{\kappa\lambda\mu\nu}$ is the Weyl
conformal tensor; $u$, $s^{\mu\nu}$, and $t^{\kappa\lambda\mu\nu}$ are
additional dynamical fields, and $V(\cdot)$ collectively denotes their
(unspecified) potentials.

In the context of SME, to be fully consistent with the Riemannian
geometric setup, the fields $u$, $s^{\mu\nu}$, and
$t^{\kappa\lambda\mu\nu}$ seek their cosmological values with
minimization of $V(\cdot)$ through the mechanism of spontaneous
symmetry breaking~\cite{kos04}. These fields acquire background values
$\bar u$, $\bar s^{\mu\nu}$, and $\bar t^{\kappa\lambda\mu\nu}$,
respectively.  Contrast to the traditional Higgs
mechanism~\cite{hig14,eng14}, the symmetry breaking here involves
tensor degrees, so the local Lorentz symmetry of spacetime is
consequently broken if the vacuum expectation values of these tensor
fields are nonzero. After adopting several plausible assumptions and
taking care of fluctuations around the vacuum expectation values,
including massless Nambu-Goldstone modes, Bailey and Kosteleck{\'y}
were able to explore the experimental phenomena of
Eq.~(\ref{eq:sme:lagrangian}) at the leading post-Newtonian
order~\cite{bk06}. At the leading order, $\bar
t^{\kappa\lambda\mu\nu}$ does not show up~\cite{bon15}, and $\bar u$
can be absorbed into a redefinition of the gravitational constant
$G$. Therefore, we are left with nine degrees of freedom with the
symmetric and traceless tensor, $\bar s^{\mu\nu}$, which is the primary object
we are to deal with in the following text. 

In an asymptotically flat coordinate frame, $\bar s^{\mu\nu}$ is simply a
constant dimensionless matrix. Between asymptotically flat frames chosen by
experimenters, $\bar s^{\mu\nu}$ transforms according to canonical Lorentz
transformation laws.\footnote{See Refs.~\cite{bk06,blu15} for distinctions
between particle Lorentz invariance and observer Lorentz invariance. In the
context of SME, particle Lorentz invariance is broken, while observer Lorentz
invariance preserves.} To facilitate comparisons between different experiments,
$\bar s^{\mu\nu}$ are often reported in the asymptotically inertial frame,
$(\partial_T,\partial_X,\partial_Y,\partial_Z)$, that is comoving with the rest
frame of the Solar System and that coincides with the canonical Sun-centered
frame~\cite{bk06}. Comparisons of SME with the PPN framework and Nordtvedt's
anisotropically parametrized post-Newtonian model~\cite{nor76} can be found in
Ref.~\cite{bk06}.

Experimental tests of the pure-gravity sector of SME include perihelion shift,
time-delay effect, lunar and satellite laser ranging, gravimeter, torsion
pendulum, gyroscope, binary pulsars, and so on~\cite{bk06}.\footnote{Recently,
  there are some experimental exploration for operators with mass dimension
  higher than four in the pure-gravity sector of SME (the so-called {\it
  non-minimal} SME). It belongs to the short-range gravity regime, and is beyond
  the scope of current paper. Interested readers are pointed to
  Refs.~\cite{bkr15,lk15,stt+15}. Moreover, phenomena with Lorentz-violating
  matter-gravity couplings can be found in Refs.~\cite{kt11,lib13,tas14,jty15}.}
  We will review results from LLR~\cite{bcs07}, atom
  interferometry~\cite{mch+08,cch+09}, Gravity Probe B~\cite{beo13}, planetary
  orbital dynamics~\cite{hbp+15}, and millisecond radio
  pulsars~\cite{xie13,shao14,shao14b}. These results are collected by the
  updating arXiv version of {\it Data tables for Lorentz and CPT
  violation}~\cite{kr11}.

\begin{table*}
  \caption{Observational constraints on the Lorentz-violating $\bar s^{\mu\nu}$
  components in the pure-gravity sector of SME~\cite{bk06} at 68\% confidence
  levels, from the combined limits of LLR and atom
  interferometry~\cite{cch+09}, Gravity Probe B~\cite{beo13}, and pulsar
  timing~\cite{shao14,shao14b}. Two limits on $\bar s^{\rm TT}$ from pulsar
  timing correspond to the cases with and without the assumption that the
  isotropic CMB frame is the preferred frame~\cite{shao14b}. See also
  Ref.~\cite{hbp+15} for a set of recent constraints from planetary orbital
  dynamics.  \label{tab:sme}}
  \begin{tabular*}{\textwidth}{lll}
    \hline\hline
    $\bar s^{\mu\nu}$ components & Pulsar timing & Other experiments\\ 
    \hline
    $|\bar s^{\rm TT}|$ & $<8\times10^{-6}$ or
    $<1.4\times10^{-4}$~\cite{shao14b} &
    $<3.8\times10^{-3}$~\cite{beo13} \\  
    $\bar s^{\rm TX}$ & $(-5.2,\,5.3)\times10^{-9}$~\cite{shao14} &
    $(-5.7,\,6.7)\times10^{-7}$~\cite{cch+09} \\ 
    $\bar s^{\rm TY}$ & $(-7.5,\,8.5)\times10^{-9}$~\cite{shao14} &
    $(-1.2,\,1.4)\times10^{-6}$~\cite{cch+09} \\ 
    $\bar s^{\rm TZ}$ & $(-5.9,\,5.8)\times10^{-9}$~\cite{shao14} &
    $(-4.2,\,3.4)\times10^{-6}$~\cite{cch+09} \\ 
    $\bar s^{\rm XY}$ & $(-3.5,\,3.6)\times10^{-11}$~\cite{shao14} &
    $(-2.1,\,0.9)\times10^{-9}$~\cite{cch+09} \\ 
    $\bar s^{\rm XZ}$ & $(-2.0,\,2.0)\times10^{-11}$~\cite{shao14} &
    $(-4.1,\,-1.3)\times10^{-9}$~\cite{cch+09} \\ 
    $\bar s^{\rm YZ}$ & $(-3.3,\,3.3)\times10^{-11}$~\cite{shao14} &
    $(-0.8,\,2.0)\times10^{-9}$~\cite{cch+09} \\ 
    $\bar s^{\rm XX}-\bar s^{\rm YY}$ &
    $(-9.7,\,10.1)\times10^{-11}$~\cite{shao14} &
    $(-2.8,\,0.4)\times10^{-9}$~\cite{cch+09} \\  
    $\bar s^{\rm XX}+\bar s^{\rm YY}-2\bar s^{\rm ZZ}$ &
    $(-12.3,\,12.2)\times10^{-11}$~\cite{shao14} &
    $(-3.6,\,4.0)\times10^{-8}$~\cite{cch+09} \\  
    \hline
  \end{tabular*}
\end{table*}

{\bf Lunar Laser Ranging.} Lunar laser ranging experiment has been ongoing since
the Apollo mission put retroreflectors on the Moon surface in 1969. It measures
the Earth-Moon separation by timing the round-trip flight of light between
ranging stations on the Earth surface and retroreflectors on the Moon
surface~\cite{bcd+73}. The precision has now reached $\delta
x/d_{\oplus\leftmoon} \sim \mbox{(a few millimeters)} / \mbox{(384400 km)} \sim
\mbox{a few parts in } 10^{12}$, where $\delta x$ is the timing uncertainty in
distance and $d_{\oplus\leftmoon}$ is the average Earth-Moon separation.

Local Lorentz invariance violation would manifest as oscillatory perturbations
to the lunar orbit at specific frequencies~\cite{bk06,bcs07},
\begin{equation}
  \delta d_{\oplus\leftmoon}^{\rm SME}(t) = \sum_n \left[
    A_n \cos(\omega_n t + \phi_n) +
    B_n \sin(\omega_n t + \phi_n) \right] \,.
\end{equation}
The dominant perturbations occur at four frequencies, $\omega_{\leftmoon}$,
$2\omega_{\leftmoon}$, $2\omega_{\leftmoon} - \omega_{\leftmoon}^\prime$, and
$\omega_\oplus$, where $\omega_{\leftmoon}$ is the sidereal lunar orbital
frequency, $\omega_{\leftmoon}^\prime$ is the anomalistic lunar orbital
frequency, and $\omega_\oplus$ is the sidereal Earth orbital frequency. The
amplitudes, $A_n$ and $B_n$, and phases, $\phi_n$, were calculated by Bailey and
Kosteleck{\'y}~\cite{bk06}. The dominant contributions to $\delta
d_{\oplus\leftmoon}^{\rm SME}(t)$ are controlled by six linear combinations of
$\bar s^{\mu\nu}$ in $A_n$ and $B_n$.

Battat et al. used 14401 normal points taken at the McDonald Laser-Ranging
Station in Texas and the Observatoire de la C\^ote d'Azur Station in Grass,
spanning from 1969 to 2003, to look for possible hints of nonzero SME
parameters~\cite{bcs07}. They found six linear combinations of $\bar s^{\mu\nu}$
consistent with zero and constrained them to the level of $10^{-6}$ to
$10^{-11}$ (see Table 3 in Ref.~\cite{bcs07}).

{\bf Atom Interferometry.} A relatively recent setup for precision tests of
gravity uses atom interferometry, where a freely falling frame is realized with
neutral atoms to an outstanding accuracy~\cite{kc91,chu02}. The quantum phase of
cold atoms is influenced by the local gravitational acceleration ${\bf g}$. The
phase difference in two interferometric paths can be measured with a great
precision when the paths are recombined at a final beam
splitter~\cite{mch+08,cch+09}.

With a hypothetical violation in local Lorentz invariance, the gravitational
acceleration develops modulations due to the rotation of the Earth and the
revolving of the Earth around the Sun,
\begin{equation}
  \left[\frac{\delta g(t)}{g_0}\right]^{\rm SME} = \sum_m \left[
    C_m \cos(\omega_m t + \psi_m) + D_m \sin(\omega_m t + \psi_m)
    \right] \,.
\end{equation}
The dominant contributions occur at frequencies $\Omega_\oplus$,
$2\Omega_\oplus$, $\Omega_\oplus \pm \omega_\oplus$, and $2\Omega_\oplus \pm
\omega_\oplus$, where $\Omega_\oplus \simeq 2\pi / (\mbox{23.93 hr})$ is the
rotation frequency of the Earth.  The amplitudes, $C_n$ and $D_n$, and phases,
$\psi_n$, can be found in Refs.~\cite{bk06,mch+08,cch+09}.

M\"uller et al.~\cite{mch+08} and Chung et al.~\cite{cch+09} used a vertical
Mach-Zehnder atom interferometer with a resolution up to $8 \times 10^{-9} g /
\sqrt{\rm Hz}$. They obtained limits on seven linear combinations of $\bar
s^{\mu\nu}$ with the SME parameters from the Lorentz-violating electromagnetic
sector, $\tilde \kappa_{e^-}$ and $\tilde \kappa_{o^+}$~\cite{ck97,ck98}. These
limits are remarkable because they probe gravitational effects at the interface
of the quantum world. After assuming vanishingly smallness of $\tilde
\kappa_{e^-}$ and $\tilde \kappa_{o^+}$~\cite{kr11}, these limits were combined
with that from LLR~\cite{bcs07}. This helps in breaking some
degeneracy of parameters, and results in tight constraints on eight components
of $\bar s^{\mu\nu}$, where one roughly has $|\bar s^{\rm TK}| \lesssim
10^{-6}$--$10^{-7}$ and $|\bar s^{\rm JK}| \lesssim 10^{-9}$ with $J, K \in \{X,
Y, Z\}$; see the third column of Table~\ref{tab:sme} for numbers.

{\bf Gravity Probe B.} With precision measurements of LLR and
atom interferometry, eight components of $\bar s^{\mu\nu}$ have been limited,
leaving only the time-time component, $\bar s^{\rm TT}$, unconstrained. This
component does not enter the orbital dynamics of the Moon nor the phase
evolution of cold atoms, hence cannot be studied with these
experiments~\cite{bk06}. Gravity Probe B is a gyroscope experiment in orbit
around the Earth, that monitors the spin dynamics of four cryogenic gyroscopes
onboard, relative to a remote guiding star. The analysis revealed a
geodetic-precessing drift and a frame-dragging drift at precision of $0.3\%$ and
$19\%$, respectively, in agreement with GR~\cite{edp+11,ove15}.

A hypothetical existence of Lorentz violation would modify the orbital dynamics
of the satellite and the spin-orbit dynamics of gyroscopes. Most notably, the
$\bar s^{\rm TT}$ component enters the equations of geodetic precession and
frame dragging, hence can be studied with Gravity Probe B~\cite{bk06}. The
precession rate of gyroscopes onboard Gravity Probe B is,
\begin{equation}
{\bf\Omega}_{\rm GPB} =
{\bf\Omega}^{\rm GR} + {\bf\Omega}^{\rm SME}\,,
\end{equation}
composing of the conventional general-relativistic contribution,
${\bf\Omega}^{\rm GR}$, and an extra contribution from SME Lorentz-violating
coefficients, ${\bf\Omega}^{\rm SME}$ (see Eqs.~(154--156) in Ref.~\cite{bk06}
for expressions).  Bailey et al.~\cite{beo13} analyzed the effects in detail and
obtained the first constraint on the time-time component, $|\bar s^{\rm TT}| <
3.8 \times 10^{-3}$, at 68\% confidence level, which was used to break all
degeneracies of previous results.

{\bf Pulsar Timing.} As in the case of LLR, a nonzero $\bar
s^{\mu\nu}$ background field would modify the orbital dynamics of two massive
bodies. For binary pulsars, because that the technique of pulsar timing has the
ability to precisely map the (projected) relativistic pulsar trajectory around
the common center of the binary motion, the modification would manifest itself
in the arrival times of pulse signals. If the modification is significantly
larger than the uncertainty of arrival times, one would see significant timing
residuals after fitting a canonical timing model based on GR to
the dataset~\cite{dd86,dt92}.

In the pure-gravity sector of SME, Lorentz invariance violation causes secular
changes in orbital elements after averaging over one orbital period~\cite{bk06},
\begin{eqnarray}
  \delta \left\langle \frac{{\rm d} \omega}{{\rm d} t}\right\rangle  &=&
  f_{\dot\omega}^{\rm SME}(P_b, e, i, \Omega, \omega, m_1, m_2; \bar s^{\mu\nu})
  \,, \label{eq:sme:omdot} \\
  \left\langle \frac{{\rm d} e}{{\rm d} t}\right\rangle &=& f_{\dot
    e}^{\rm SME}(P_b, e, i, \Omega, \omega, m_1, m_2; \bar s^{\mu\nu})
  \,, \label{eq:sme:edot}\\
  \left\langle \frac{{\rm d} x}{{\rm d} t}\right\rangle &=& f_{\dot
    x}^{\rm SME}(P_b, e, i, \Omega, \omega, m_1, m_2; \bar s^{\mu\nu})
  \,, \label{eq:sme:xdot}
\end{eqnarray}
where $\delta \left\langle {{\rm d} \omega}/{{\rm d} t}\right\rangle$ is the
difference in periastron advance rate with respect to  GR, while
the contributions from GR to $\left\langle {{\rm d} e}/{{\rm d}
t}\right\rangle$ and $\left\langle {{\rm d} x}/{{\rm d} t}\right\rangle$ are too
small compared with current timing precision~\cite{lk04}.  Functions
$f_{\dot\omega}^{\rm SME}(\cdots;\cdot)$, $f_{\dot e}^{\rm SME}(\cdots;\cdot)$,
and $f_{\dot x}^{\rm SME}(\cdots;\cdot)$ are homogeneous linear forms of $\bar
s^{\mu\nu}$, and their explicit expressions can be found in
Refs.~\cite{bk06,shao14,shao14b}. It is important to note that these functions
depend on the projected $\bar s^{\mu\nu}$ onto the coordinate system defined by
the orbit (see Figure 2 in Ref.~\cite{shao14} for the definition of the
coordinate system). The projected $\bar s^{\mu\nu}$ can be related to its values
in the standard frame of SME through angles ($\alpha$, $\delta$) and ($i$,
$\Omega$, $\omega$). The rotation matrices of transformation were given
explicitly in the supplemental material of Ref.~\cite{shao14} and in
Eqs.~(19--24) of Ref.~\cite{shao14b}. While the underlying $\bar s^{\mu\nu}$
field is the same for all binary pulsars, because of different sky locations,
characterized by ($\alpha$, $\delta$), different orbital orientations,
characterized by ($i$, $\Omega$, $\omega$), different orbital characteristics
(e.g., orbital period, $P_b$, orbital eccentricity, $e$, and so on), the
projections are different for different binary pulsars.  Therefore, a variety of
binary pulsars probe a variety of linear combinations of $\bar s^{\mu\nu}$. In
the area of pulsar astronomy, the advanced radio telescopes have provided us
with dozens of well-timed millisecond binary pulsars (especially within the
programs of pulsar timing array~\cite{hob13,mcl13,kc13,man13}). It becomes
possible to use an array of pulsars to constrain different linear combinations
of $\bar s^{\mu\nu}$, and finally put all together to break all degeneracy of
underlying degrees of freedom with negligible mutual correlation.

\begin{figure*}
  \centering
  \includegraphics[width=12cm]{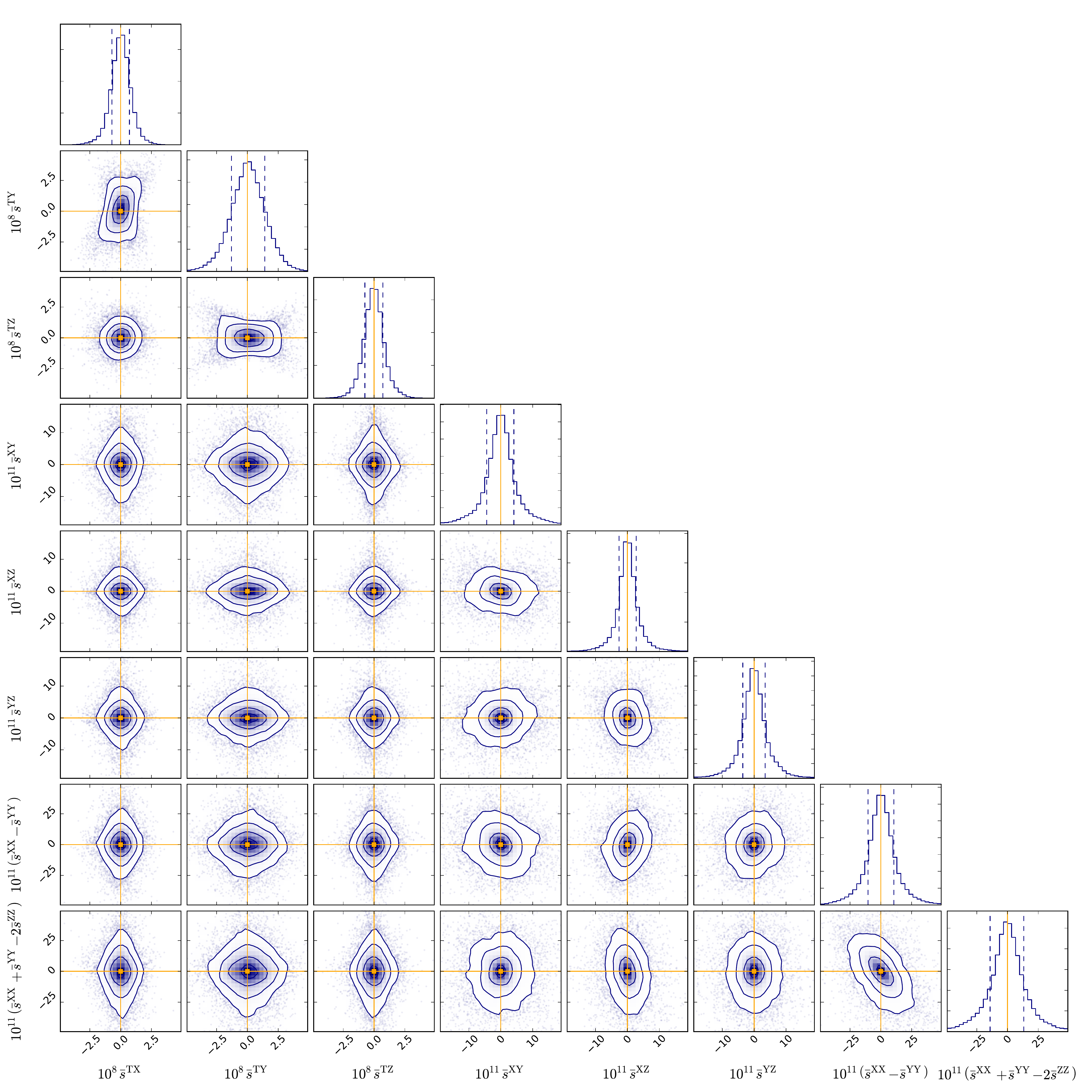}
  \caption{\label{fig:sme} Components of $\bar s^{\mu\nu}$ that pass all the
  tests in the simulation. Their marginalized probability densities are depicted
  as histograms where dash lines enclose the 68\% confidence level. Values of
  $\bar s^{\mu\nu}$ in GR are shown in orange. Data are taken from
  Ref.~\cite{shao14}.}
\end{figure*}

However, there are some subtleties related to observational characteristics of
binary pulsars~\cite{shao14}:
\begin{enumerate}
\item Concerning Eqs.~(\ref{eq:sme:edot}--\ref{eq:sme:xdot}), usually, $\dot e$
  and $\dot x$ are not reported in literature.  Conservative surrogates of their
  values used the observed uncertainties of $e$ and $x$, by assuming that the
  observed uncertainties are purely generated from the Lorentz-violating
  linear-in-time evolution. If there are other contributions to these
  quantities, the uncertainties are likely to be larger.
\item Because in most binary pulsars, the determination of component masses
  heavily rely on the measured value of $\dot\omega$. Therefore, to avoid
  improper cyclic usage, the $\dot\omega$ measurement must not be used in
  Eq.~(\ref{eq:sme:omdot}) to test local Lorentz invariance violation, unless
  the masses can be determined from other observations without using
$\dot\omega$. In the sample pulsars of
  Ref.~\cite{shao14}, three white dwarf -- neutron star binaries have
  double-line observations with radio timing on the pulsars and optical
  spectroscopy on the white dwarfs. The masses of these binaries were determined
  accurately with the mass ratio from orbital-phase-resolved radio and optical
  data, and well-tested white dwarf atmosphere
  models~\cite{lwj+09,fwe+12,afw+13}. For consistency, the test in
  Eq.~(\ref{eq:sme:omdot}) were performed only for these three binaries.
\item A large $\dot\omega$ can render the secular changes nonlinear in
  Eqs.~(\ref{eq:sme:omdot}--\ref{eq:sme:xdot})~\cite{wk07}.
  Fortunately, the linear-in-time approximation is plausible with the binary
  samples in Ref.~\cite{shao14} at current stage. Otherwise, more proper timing
  models are needed; see Ref.~\cite{wk07} for example.
\item To precisely project $\bar s^{\mu\nu}$ on the binary orbital frame, one
  has to account for the boost between the pulsar frame and the frame of the
  Solar System. However, the 3-dimensional bulky velocity of binaries, ${\bf
  v}^{\rm binary}$, is not always observationally known. From our understanding
  of binary evolution, the velocity is usually small, $|{\bf v}^{\rm
  binary}|\sim 10^2 \mbox{ km s}^{-1}$, compared with the light speed, $c = 3.0
  \times 10^5 \mbox{ km s}^{-1}$. So at the first-order approximation, the boost
  effect can be neglected~\cite{shao14}. More on how to wisely use this boost
  to probe the $\bar s^{\rm TT}$ component will be discussed later~\cite{shao14b}.
\item To fully determine the projection of $\bar s^{\mu\nu}$ mentioned before,
  one has to know the absolute orientation of binary orbits, characterized by
  $(i,\Omega,\omega)$, wherein the longitude of ascending node, $\Omega$, is not
  observationally available for most binary pulsars. Therefore, a probabilistic
  assumption of $\Omega$ was made.
\end{enumerate}

In addition to the modifications in the orbital dynamics of binary pulsars,
Eqs.~(\ref{eq:sme:omdot}--\ref{eq:sme:xdot}), a nonzero $\bar s^{\mu\nu}$ field
would also cause a free precession of the spin of a solitary
pulsar~\cite{shao14} at a drift rate,
\begin{equation}
  \Omega_k^{\rm SME} = \sum_j \frac{\pi}{P} \bar s^{jk} \hat{S}^j \,,
\end{equation}
where $P$ is the spin period, and $\hat{\bf S}$ is a unit vector pointing in the
direction of the pulsar spin. Again, the $\bar s^{\mu\nu}$ field is projected in
the frame established by the pulsar system. Such a spin precession would change
our line-of-sight cut on the pulsar radiation zone, consequently resulting in
changes in the pulse profile as a function of time. It is similar to the case of
a nonzero PPN $\alpha_2$. Null observation of such changes can be used to
constrain SME parameters~\cite{sck+13} (see Figure~\ref{fig:B1937} for an
illustration on the stability of pulse profiles for millisecond pulsars). The
azimuthal angle of the pulsar spin is in general not observable, therefore a
similar probabilistic consideration as for the $\Omega$ in binary pulsars was
carried out~\cite{shao14}.

Combining all concerns above for binary pulsars and solitary pulsars, Monte
Carlo simulations were set up to perform simultaneously twenty-seven tests from
thirteen pulsar systems. Because in total there are eight degrees of freedom
entering the orbital dynamics of binary pulsars and spin precession of solitary
pulsars, twenty-seven tests have over-constrained them. For a set of $\bar
s^{\mu\nu}$ to be consistent with observations, all twenty-seven tests have to
be passed within observational uncertainties. This reduces correlations between
the components of $\bar s^{\mu\nu}$ numerously. Figure~\ref{fig:sme} illustrated
the values and distributions of $\bar s^{\mu\nu}$ that pass all tests in the
simulation~\cite{shao14}; see the second column in Table~\ref{tab:sme} for
statistical numbers. These limits are very tight limits in the experimental
post-Newtonian gravity~\cite{kr11}.

As in the cases of LLR and atom interferometry, the $\bar s^{\rm
TT}$ component does not enter the orbital dynamics of binary pulsars and spin
precession of solitary pulsars, if the boost between different frames is
neglected~\cite{bk06,beo13,shao14b}.  An idea of using the boost effect between
different Lorentz frames was proposed to constrain $\bar s^{\rm TT}$ in the
context of binary pulsar timing~\cite{shao14b}. A full Lorentz transformation
includes rotations and boosts.  When the frame comoving with the binary has a
relative velocity with respect to the standard frame of SME, the boost will mix
the $\bar s^{\rm TT}$ component in the standard frame into the other components
of $\bar s^{\mu\nu}$ in the binary-comoving frame~\cite{bk06,shao14b}.
Therefore, through constraining the time-spatial and spatial-spatial components
in the binary-comoving frame, one can effectively constrain the $\bar s^{\rm
TT}$ component in the standard SME frame.

In order to use the boost, one needs to have the knowledge of 3-dimensional
bulky velocity of the binary with respect to the Solar System. This is possible
for the aforementioned three white dwarf -- neutron star binaries with both
radio and optical observations~\cite{lwj+09,fwe+12,afw+13}. If one focuses on
the possibility that there exists a preferred frame in SME, then the only
nonvanishing component of $\bar s^{\mu\nu}$ is $\bar s^{\rm TT}$ in this
preferred frame. In this case it is possible to directly constrain $\bar s^{\rm
TT}$ even without knowing the longitude of ascending node, $\Omega$. This was
done by obtaining the limit of $\bar s^{\rm TT}$ as a function of $\Omega$, and
then conservatively choosing the worst one, similar to the direct test of
$\alpha_1$~\cite{sw12}.\footnote{When $\bar s^{\mu\nu} = 0$ except $\bar s^{\rm
TT} = \bar s^{\rm XX} + \bar s^{\rm YY} + \bar s^{\rm ZZ}$, there are three
tests, namely Eqs.~(\ref{eq:sme:omdot}--\ref{eq:sme:xdot}), that have to be
satisfied for each binary pulsar. These tests are proved to compensate mutually
and give decent constraints for every $\Omega$. See Figures 2--4 in
Ref.~\cite{shao14b} for details.}

\begin{figure}[H]
  \centering
  \includegraphics[width=9cm]{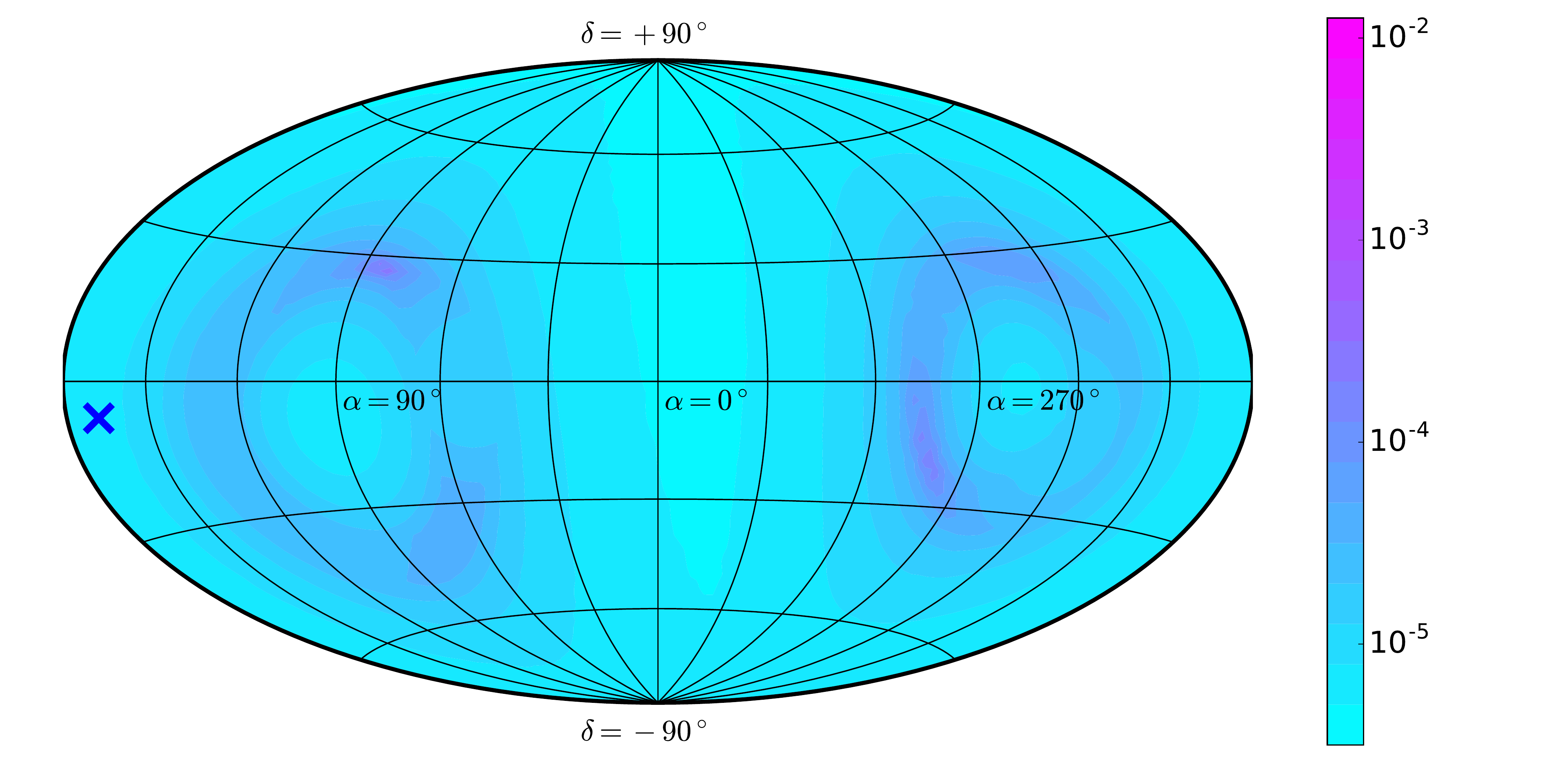}
  \caption{\label{fig:stt} Combined limits on $|\bar s^{\rm TT}|$ from three
  binary pulsars for different preferred frames. Here $(\alpha,\delta)$ denotes
  the movement direction of the Solar System with respect to a preferred frame.
  The magnitude of its velocity is assumed to be $369 \mbox{ km s}^{-1}$. The
  blue cross is the direction of the frame where CMB is isotropic at first-order
  approximation (the CMB frame). Data are taken from Ref.~\cite{shao14b}.}
\end{figure}

PSR~J1738+0333~\cite{fwe+12} turns out to be the best system to constrain the
local Lorentz invariance in gravity~\cite{sw12,shao14b}.  If one assumes that
the CMB defines a preferred frame, then calculations gave $|\bar s^{\rm TT}| < 8
\times 10^{-6}$ at 68\% confidence level~\cite{shao14b}. More generally, without
assuming the existence of a preferred frame, one gets $|\bar s^{\rm TT}| < 1.4
\times 10^{-4}$ at 68\% confidence level~\cite{shao14b}.  

If there exists a preferred frame that is different from the CMB frame, binary
pulsars are still valuable in constraining SME.  Figure~\ref{fig:stt}
illustrates the concept that, because of multiple pulsar systems with different
orientations, it is possible to constrain any kind of preferred-frame effects.
The figure shows the combined limits that are obtained from PSRs~J1738+0333,
J1012+5307, and J0348+0432 for any direction of the preferred
frame~\cite{shao14b}. It is clear that with three binaries, whatever direction
is the preferred frame, the local Lorentz invariance violation in SME can be
constrained to a level better than $10^{-4}$. 

Very recently, Jennings et al. included violations of local Lorentz invariance
in the gravitationally coupled matter sector and studied the possibility to
constrain these matter-gravity couplings with pulsar timing
experiments~\cite{kt11,jty15}. It is an extension of the previous study by
Bailey and Kosteleck{\'y} where only the gravity degree was
considered~\cite{bk06}. Because of the large degrees of freedom in parameter
estimation, the practical use of a complete timing formula to a specific binary
pulsar is not clear yet. A simultaneous fitting to multiple binary pulsars might
help breaking some degeneracy. 

{\bf Planetary Orbital Dynamics.} In the course of preparing this review, Hees
et al. analyzed the planetary orbital dynamics within the framework of SME. The
supplementary advances of the perihelia and of the nodes obtained by planetary
ephemerides analysis were used to constrain SME coefficients from the pure
gravity sector and also from gravity-matter couplings~\cite{hbp+15}. The
observations of Mercury, Venus, Earth-Moon barycenter,\footnote{In the case of
  Earth-Moon barycenter, a full 3-body dynamics was neglected in
  Ref.~\cite{hbp+15} without justification.} Mars, Jupiter, and Saturn in
  INPOP10a were nicely utilized to constrain different combinations of $\bar
  s^{\mu\nu}$ and gravity-matter couplings. However, because of the small
  inclinations of planetary orbits with respect to each other, the limits on
  $\bar s^{\mu\nu}$ are highly correlated (see their Table II). After combining
  their limits with that of LLR and atom interferometry, the
  correlations can be vastly broken~\cite{hbp+15}, and results comparable with
  that of pulsar timing were obtained.

In Section~\ref{sec:lli:ppn} and Section~\ref{sec:lli:sme}, only the
conservative dynamics with local Lorentz invariance violation were considered.
The dissipative effects, associated with gravitational radiation, can also be
used to constrain local Lorentz invariance in gravity effectively; as mentioned
in Section~\ref{sec:uff}. Yagi et al.  analyzed in detail the modification of
orbital decay rate (notably with gravitational dipole radiation and quadrupole
component of the emission), caused by a preferred time direction, in the
khronometric theory (a possible completion is Ho{\v r}ava-Lifshitz gravity) and
Einstein-\AE{}ther theory~\cite{ybyb14,ybby14,ybby14e}.  They found that the
orbital evolution of binary pulsars depends critically on the sensitivities of
the stars, which measure how their binding energies depend on the motion
relative to the preferred frame. With precision measurements of orbital decay
from PSRs J1141$-$6545, J0348+0432, J0737$-$3039, and J1738+0333, they were able
to constrain the parameters in above two theories tightly (see e.g., Figure 2 in
Ref.~\cite{ybyb14}). Thus it extends previous tests of dipole radiation in
Lorentz-invariant theories~\cite{dt92,de93} (see Section~\ref{sec:uff}) to
Lorentz-violating theories.

\section{Translational Symmetries and Conservation of Energy-Momentum}
\label{sec:trans}

In this section, we will briefly discuss experimental constraints on the
conservations of energy and momentum in gravity theories. 
\begin{itemize}
  \item In the SME framework, as the theory is defined by an invariant action
    principle, see Eq.~(\ref{eq:sme:lagrangian}), that includes no explicit
    spacetime dependence, translational symmetries are preserved. According to
    Noether's theorem, energy and momentum are conserved in SME~\cite{tll73}.
  \item In the PPN framework, the violation of conservation of total momentum 
    is described by
    five PPN parameters, $\alpha_3$ and $\zeta_i$ ($i=1,2,3,4$).  If the action
    and reaction of gravitational force are equal (Newton's third law for the
    gravitational interaction), the momentum of the system should be conserved
    in the post-Newtonian limit. Therefore, the possible nonzero values of
    $\alpha_3$ and $\zeta_i$ could imply a possible difference in the active
    gravitational mass (that produces the ``action'') and the passive
    gravitational mass (that receives the ``action''). Therefore, we will
    discuss tests of momentum-conservation-violating parameters in the PPN
    formalism.
\end{itemize}

Among these nonconservative PPN parameters, the best constrained one is
$\alpha_3$. The self-acceleration effect of $\alpha_3$ on binary pulsars that
leads to polarization of orbit was discussed in Section~\ref{sec:lli:ppn}.
An extra acceleration, ${\bf a}_{\alpha_3} \propto {\bf v}^0 \times {\bf \Omega}$,
see Eq.~(\ref{eq:ppn:a3}), that is perpendicular to the plane defined by pulsar's
absolute velocity, ${\bf v}^0$, and spin, is constrained, and leads to a limit
of $|\alpha_3|$ at the level of ${\cal O}(10^{-20})$.

In the set of $\zeta_i$'s, $\zeta_1$ was constrained indirectly through the
Nordtvedt effect that describes the violation of SEP (see
Section~\ref{sec:uff}).  The Nordtvedt parameter, $\eta_N = (m_I - m_P) / E_{\rm
grav}$, describes the difference between the inertial mass, $m_I$,
and the passive gravitational mass, $m_P$, normalized to the gravitational
self-energy of the body, $E_{\rm grav}$. $\zeta_1$ enters the Nordtvedt
parameter together with other PPN parameters,
\begin{equation}
  \eta_N = 4\beta - \gamma -3 - \frac{10}{3}\xi - \alpha_1 +
  \frac{2}{3}\alpha_2 - \frac{2}{3}\zeta_1 - \frac{1}{3}\zeta_2 \,.
  \label{eq:nordtvedt:par}
\end{equation}
When $\beta$, $\gamma$, $\xi$, $\alpha_1$, and $\zeta_2$ are constrained
independently, a limit of $\eta_N$ could lead to a constraint on $\zeta_1$.  As
mentioned in Section~\ref{sec:uff}, the best limit of $\eta_N$ comes from LLR 
by constraining  possible perturbation in the Earth-Moon
distance. Because of the improvements in other PPN parameters in
Eq.~(\ref{eq:nordtvedt:par}) with new experiments, the precision of $\zeta_1$ is
currently limited by the precision of $\eta_N$, and one can infer $|\zeta_1|
\lesssim 10^{-3}$. 

The PPN parameter $\zeta_2$ measures how much gravity is produced by the 
gravitational energy~\cite{will92}.  Together with $\alpha_3$, it produces an 
extra self-acceleration in direction to the periastron of a binary orbit. As a
consequence, the acceleration introduces a nonvanishing value for the second
time derivative of the orbital period. With the observational upper limit from
PSR~B1913+16, Will was able to constrain $|\alpha_3 + \zeta_2| < 4 \times
10^{-5}$~\cite{will92}.  Because $\alpha_3$ is constrained with much more
sensitive test independently (see Section~\ref{sec:lli:ppn}), the above limit
reduces to $|\zeta_2| < 4\times10^{-5}$.

The PPN parameter $\zeta_3$ produces an asymmetry between the passive
gravitational mass, $m_P$, and the active gravitational mass, $m_A$, as $m_A -
m_P = \frac{1}{3} \zeta_3 E_{\rm EM,static}$, where $E_{\rm EM,static}$ is the
electrostatic binding energy of the body. With the Moon and the Earth well
modeled, one can convert the limit of anomalous secular acceleration in the
Sun-Earth-Moon dynamics from LLR into a limit of $\zeta_3$. By
this, $|\zeta_3|$ is constrained to be less than $10^{-8}$~\cite{bb86}. One 
needs to keep in mind that the limit of $\zeta_3$ could have degeneracy with a 
nonzero $\dot{G}/G$~\cite{will14} (see next section).

There is no direct constraint yet on the last parameter, $\zeta_4$.
Nevertheless, there is strong theoretical evidence that this parameter is
dependent on other PPN parameters~\cite{will76},
\begin{equation}
  \zeta_4 = \frac{1}{2} \alpha_3 + \frac{1}{3} \zeta_1 - \frac{1}{2} \zeta_3 \,.
  \label{eq:zeta4}
\end{equation}
Such a dependence is expected because of an expected relation of the
gravitational effects from kinetic energy, internal energy, and pressure of
matter~\cite{will14}. It is possible to estimate a limit on $\zeta_4$ at the
level of ${\cal O}(10^{-3})$ from the limits of $\alpha_3$, $\zeta_1$, and
$\zeta_3$ with above relation.

To shortly summarize, GR predicts that, for an isolated gravitating system in
asymptotically flat spacetime, energy and momentum are
conserved\footnote{With gravitational radiation, one needs to add up the amount
of energy and momentum at infinity, namely, gravitational waves, through the
balance equation.}~\cite{mtw73}.  Various experiments proved that in the
gravitational interaction, energy and momentum are conserved to a very high
precision, consistent with the prediction of GR.

\section{Time Varying Gravitational Constant}
\label{sec:Gdot}

The locally measured Newtonian gravitational constant $G$ may vary with time as
the Universe expands. In fact, this is expected for most alternatives to GR that
violate SEP\cite{will93}. For the $T_1(\alpha_0,\beta_0)$ class of scalar-tensor
theories, introduced in Section~\ref{sec:uff}, the temporal change in $G$ can be
expressed in terms of the time varying asymptotic scalar field
$\varphi_\infty(t)$ of a local system.  To linear order one finds \cite{uzan11},
\begin{equation}\label{eq:GdotT1}
   \frac{\dot G}{G} = 2\left[1 + \frac{\beta_0}{1 + \alpha_0^2}\right]
                      \alpha_0\dot\varphi_\infty 
                    = \eta_{\rm N} \,\frac{1 + \alpha_0^2}{\alpha_0} \,
                      \dot\varphi_\infty 
                    \simeq \eta_{\rm N} \, \frac{\dot\varphi_\infty}{\alpha_0}\;,
\end{equation}
where in the last step we have used the approximation $(1 + \alpha_0^2)/\alpha_0
\simeq 1/\alpha_0$, a consequence of the experimental evidence that $\alpha_0^2
\ll 1$. The introduction of the Nordtvedt
parameter $\eta_{\rm N}$ in Eq.~(\ref{eq:GdotT1}) illustrates nicely the
connection between $\dot G$ and a violation of SEP. The quantity $\dot\varphi_0$
can be calculated from a cosmological model, and by this be linked to the Hubble
constant $H_0$ (see for instance, Ref.~\cite{wein72}). More generally, one can
write for the present change of $G$,
\begin{equation}
  \frac{\dot G}{G} = \sigma \, H_0 \;,
\end{equation}
where the value of the Hubble constant can be taken from Ref.~\cite{PC13},  $H_0 =
67.8\,{\rm km\,s}^{-1}{\rm \,Mpc}^{-1}$.

By now there are various experiments that tightly constrain a change in the
gravitational constant on different time scales. Some of them probe a change
over cosmological time scales, i.e.~$G(t)$, others constrain a change of today's
$\dot{G}$ (see Ref.~\cite{uzan11} for a review). 

In a binary system, a time variation of $G$ changes the orbital period $P_b$.
If the gravitational binding energy of the masses is small, like for Solar
system bodies, the change of the orbital period is to first order given by
\cite{dgt88},
\begin{equation}\label{eq:PdotGdot}
  \frac{\dot{P}_b}{P_b} = -\frac{\dot{n}_b}{n_b} = -2 \, \frac{\dot{G}}{G} \;,
\end{equation}
and the semimajor axis of the relative motion changes according to,
\begin{equation}
  \frac{\dot{a}}{a} = -\frac{\dot{G}}{G} \;.
\end{equation}
In the Solar system, LLR gives \cite{hmb10} 
\begin{eqnarray}\label{eq:GdotLimitLLR}
  \frac{\dot{G}}{G} &=& (-0.7 \pm 3.8) \times 10^{-13} \, {\rm yr}^{-1} 
  					\nonumber\\
                    &=& (-0.0010 \pm 0.0055) \, H_0 \;,
\end{eqnarray}
and from planetary ephemerides one has \cite{fle+14}
\begin{eqnarray}\label{eq:GdotLimitPl}
  \frac{\dot{G}}{G} &=& (0.1 \pm 1.8) \times 10^{-13} \, {\rm yr}^{-1} 
  					\nonumber\\
                    &=& (0.0001 \pm 0.0026) \, H_0 \;.
\end{eqnarray}

When one uses binary pulsars to test for a time varying gravitational constant,
one has to keep in mind that Eq.~(\ref{eq:PdotGdot}) is not applicable to
systems containing strongly self-gravitating bodies \cite{nor90}. This can be
made plausible by the following argument. A change in $G$ changes the
gravitational binding energy of a self-gravitating body, and by this its mass.
While such a change is negligible in the Earth-Moon system, since the fractional
binding energy is very small for these bodies ($\epsilon_{\rm Earth} \approx -5
\times 10^{-10}$), it is significant for neutron stars, where the gravitational
self-energy accounts for a significant fraction of the total mass
($\epsilon_{\rm NS} \sim -0.1$). A detailed calculation can be found in
Ref.~\cite{nor90}, which shows that for a binary pulsar system
Eq.~(\ref{eq:PdotGdot}) has to be modified to 
\begin{equation}\label{eq:PdotGdotPsr}
  \frac{\dot{P}_b}{P_b} = -2 \, \frac{\dot{G}}{G} 
  \left[1 - \left(1 + \frac{m_c}{2M}\right) s_p 
          - \left(1 + \frac{m_p}{2M}\right) s_c \right] \;,
\end{equation}
where $m_p$ and $m_c$ denote pulsar and companion mass respectively, and $M =
m_p + m_c$. The ``sensitivity'',
\begin{equation}
  s_A \equiv -\left.\frac{\partial(\ln m_A)}{\partial(\ln G)}\right|_N \,,
\end{equation}
measures how the mass of body $A$ changes with a change of the local
gravitational constant $G$, for a fixed baryon number $N$ (see
Ref.~\cite{will93} for details). For a given mass, the sensitivity of a neutron
star depends on the equation of state and on the specifics of the gravity
theory. If the companion of the pulsar is a weakly self-gravitating star, like a
white dwarf, $s_c$ becomes negligible and Eq.~(\ref{eq:PdotGdotPsr}) simplifies
to
\begin{equation}\label{eq:PdotGdotPsrWD}
  \frac{\dot{P}_b}{P_b} \simeq -2 \, \frac{\dot{G}}{G} 
  \left[1 - \left(1 + \frac{m_c}{2M}\right) s_p \right] \;.
\end{equation}

The currently best pulsar limit for a change in the gravitational constant comes
from the pulsar -- white dwarf system, PSR~J1713$+$0747. PSR~J1713$+$0747 is a
1.3\,$M_\odot$ millisecond pulsar, which is in a wide ($P_b = 67.8$\,d) low
eccentricity orbit with a 0.29\,$M_\odot$ white dwarf companion. Based on 21
years of timing data for that system, Zhu et al.~\cite{zsd+15} derived, 
\begin{eqnarray}\label{eq:GdotLimit1713}
  \frac{\dot{G}}{G} &=& (-6 \pm 11) \times 10^{-13} \, {\rm yr}^{-1} 
  					\nonumber\\
                    &=& (0.009 \pm 0.016) \, H_0\;,
\end{eqnarray}
with 95\% confidence. It is generally expected that a gravity theory with a
varying gravitational constant also predicts the existence of dipolar
gravitational waves, a further consequence of SEP violation, as discussed in
Section~\ref{sec:uff}. The emission of dipolar radiation modifies $\dot{P}_b$,
and could in principle even balance a significant part of a change in $P_b$
caused by $\dot G$. In the absence of non-perturbative strong-field effects one
finds for the change in the orbital period of a pulsar -- white dwarf system in
a small-eccentricity ($e \ll 1$) orbit the combined expression (see
Ref.~\cite{lwj+09}),
\begin{eqnarray}
  \frac{\dot{P}_b - \dot{P}_b^{\rm GR}}{P_b} 
  &\simeq& -2 \, \frac{\dot{G}}{G}
                 \left[1 - \left(1 + \frac{m_c}{2M}\right) s_p \right] 
           \nonumber\\
  &&       - \frac{4\pi^2}{P_b^2} \, \frac{Gm_pm_c}{c^3M} \, \kappa_D s_p^2 
           \quad + {\cal O}(s_p^3) \;.
\end{eqnarray}
The constant $\kappa_D$ is a theory dependent constant, which is a priori
unknown in a generic test, where no specific gravity theory is applied. As
proposed in Ref.~\cite{lwj+09}, it is then possible to combine two pulsars with
a sufficiently large difference in their orbital periods, $P_b$, to constrain
$\dot{G}$ and $\kappa_D$ simultaneously. In Ref.~\cite{zsd+15}, Zhu et al.~have
used PSRs J1012+5307 and J1738+0333, in combination with PSR~J1713+0747 for
their analysis. This way they determined, besides the result
(\ref{eq:GdotLimit1713}), the constraint $\kappa_D = (-0.9 \pm 3.3) \times
10^{-4}$ (95\% confidence). As a final note, in such a generic test, as outlined
above, one has to make certain reasonable assumptions about the value of $s_A$
and how it changes with the pulsar mass, since the three pulsars used in
Ref.~\cite{zsd+15} have quite different masses. Certain aspects of strong-field
gravity related to $\dot G$ cannot be captured within this framework, as we will
discuss below.

As one can see from Eq.~(\ref{eq:GdotLimit1713}), the pulsar limit on $\dot{G}$
is still somewhat weaker than the ones from the Solar system; see
Eqs.~(\ref{eq:GdotLimitLLR}) and (\ref{eq:GdotLimitPl}). However, binary pulsar
experiments are sensitive to strong-field related aspects of $\dot G$, which can
in principle lead to an enhancement of $\dot{G}$, an effect that is not captured
by the linear-in-$s_A$ expression (\ref{eq:PdotGdotPsrWD}). This has been
demonstrated in Ref.~\cite{wex14}, within the  $T_1(\alpha_0,\beta_0)$ class of
mono-scalar-tensor theories \cite{de93,de96a}. Most easily this is seen, when
looking at the effective gravitational constant ${\cal G}_{AB}$ introduced in
Section~\ref{sec:uff}. For the effective gravitational constant, describing the
gravitational interaction between two strongly self-gravitating bodies (as
measured in the physical Jordan-frame), Eq.~(\ref{eq:GdotT1}) changes to,
\begin{equation}\label{eq:GdotT1_PSR}
  \frac{	\dot{\cal G}_{12}}{{\cal G}_{12}} = 
    2\left[1 +
    \frac{(\sigma_1/\alpha_0)\beta_2 + (\sigma_2/\alpha_0)\beta_1}
         {2(1 + \sigma_1\sigma_2)}\right]
    \alpha_0\dot{\varphi}_0 \;,
\end{equation}
where $\beta_A = \partial\sigma_A/\partial\varphi_\infty$. For two weakly
self-gravitating masses, Eq.~(\ref{eq:GdotT1_PSR}) approaches
Eq.~(\ref{eq:GdotT1}), since $\sigma_A/\alpha_0 \rightarrow 1$ and $\beta_A
\rightarrow \beta_0$. On the other hand, in the presence of significant
scalarization effects in the strong gravitational fields of neutron stars, the
expression in brackets of Eq.~(\ref{eq:GdotT1_PSR}) can be considerably larger
than the corresponding expression in Eq.~(\ref{eq:GdotT1}), even for $\beta_0$
values which are not yet excluded by binary pulsar experiments (see
Ref.~\cite{wex14}). As a conclusion, $\dot{G}$ tests with binary pulsars can be
more sensitive than Solar system tests in situations where a change in the
gravitational constant gets enhanced by strongly non-linear strong-field effects
in neutron stars. The details depend on the specifics of the gravity theory and
the mass of the neutron star.

\section{Summary}
\label{sec:sum}

At the time when Nordtvedt and Will built the parametrized post-Newtonian (PPN)
formalism for post-Newtonian gravity, more than forty years ago~\cite{wn72}, the
WEP was very well examined experimentally. Position invariance and Lorentz
invariance for non-gravitational physics were also well established.  These
empirical facts, that constitute the Einstein equivalence principle (EEP), are the
foundation for the PPN formalism~\cite{will93}. In the PPN formalism,
post-Newtonian physics is required to comply with a metric theory of gravity
which is a result of EEP.  The strong equivalence
principle (SEP) can be violated within the PPN formalism when PPN parameters
deviate from their GR predictions.  In the past few decades the PPN formalism
formed the main foundation for testing gravity in the Solar system and, for some
of its aspect, also for binary pulsars. Continuously accumulating data now point
towards SEP at very high precision at the first post-Newtonian order.  It
becomes even harder today for theorists to build a gravity theory beyond GR.
The new theory either needs a novel mechanism to ``screen'' deviations from GR
in aforementioned gravity experiments, or comply with the results at the probed
precision, say, by designing a gravity theory that incorporates extremely small
extra couplings or that deviates from GR beginning at the second or even higher
post-Newtonian orders. Even at higher orders, some effects are constrained as
well, e.g., the reaction of gravitational radiation on the orbital period of a
binary pulsar~\cite{wex14}.

It seems that Einstein's general relativity (GR) is the only valid gravity
theory that fully respects SEP~\cite{will14}. If one loosens some requirements
from SEP to extend the arena of gravity theories, it might be possible to build
a new gravity theory that is able to present new features with theoretical
and/or phenomenal advantages.  Nevertheless, a new gravity theory must be able
to conform with current existing experimental data that support SEP at exquisite
precision. In this short review, we presented some modern tests of important
founding principles of current wisdom in the gravity community.  Well designed
and well performed experiments from pulsar astronomy were highlighted, as they
provide very constraining results in various aspects of gravitation, and in the
meantime, encode some strong-field dynamics associated with neutron stars. For
these tests, strictly speaking, some results on tests of local position
invariance, local Lorentz invariance, and conservation laws of gravitation
represent tests of {\it effective} strong-field generalizations of the
(originally designed, weak-field, slow-motion) post-Newtonian gravity in PPN and
SME frameworks. However, previous knowledge of strong-field dynamics hints at
even tighter limits on weak-field correspondents or new limits on unprobed
higher-order coefficients if strong-field contributions are explicitly taken
into account. For example, if the $\alpha_2$ limit obtained from pulse-profile
observations of solitary pulsar~\cite{sck+13} can be expanded with (in
accordance with the expansion of the difference between inertial mass and
gravitational mass, in terms of the Nordtvedt parameter $\eta_N$),
\begin{equation}\label{eq:a2exp}
  \alpha_2 = \alpha_2^{(0)} + \alpha_2^{(1)} \frac{E_{\rm grav}}{Mc^2}  +
  \alpha_2^{(2)} \left( \frac{E_{\rm grav}}{Mc^2} \right)^2 + \cdots\,,
\end{equation}
then the limit in Eq.~(\ref{eq:a2:limit}) not only constrains the weak-field
counterpart, $\alpha_2^{(0)}$, at the level of ${\cal O}(10^{-9})$, but also
poses strong constraints on higher-order terms, $\alpha_2^{(1)}$,
$\alpha_2^{(2)}$, and so on, at the levels of ${\cal O}(10^{-8})$, ${\cal
O}(10^{-7})$, and so on, with $| E_{\rm grav} / Mc^2 | \simeq 0.1$ for compact
neutron stars, in particular when combined with Solar system constraints.
Detailed accurate mapping of strong-field generalization and weak-field
counterpart needs explicit calculations in specified gravity theories, for
example, in scalar-tensor theories of Damour and Esposito-Far{\`
e}se~\cite{de92}. Also, equations like (\ref{eq:a2exp}) completely fail to
capture non-perturbative strong field deviations, like spontaneous scalarization
\cite{de93}. To understand the meaning of pulsar limits in such situations, it
is very helpful to discuss pulsars within a theory-specific framework, like what
we have done here for the $T_1(\alpha_0,\beta_0)$ class of scalar-tensor
theories. 

Although pulsar timing allows us to test some strong-field
dynamics, in particular in the
quasi-stationary strong-field regime, experiments posing a complete set of
constraints on strong-field orbital dynamics of binaries are only at dawn.  The
upcoming gravitational wave observations from ground- and space-based detectors
will undoubtedly open the window to the highly dynamical strong-field regime
with typical orbital velocities of $v_{\rm orb} \sim 0.1\,c$. With the first
gravitational-wave event, GW150914, detected at the Laser Interferometer
Gravitational-wave Observatory (LIGO)~\cite{ligo16}, a new era for testing
gravitation in strong gravitational field has started~\cite{will14,bs14,ys13}.
Tests of strong-field gravity from gravitational waves and from pulsars are
complementary to each other. For example, in testing scalar-tensor theories,
coalescence of neutron stars and timing of binary pulsars could probe different
portion of parameter space, as well as different scenarios of scalarization
(nondynamical spontaneous scalarization~\cite{de98} versus dynamical
scalarization~\cite{bppl13,stob14}).  In the parameter space where those two kinds of
observation overlap, the results from gravitational waves and from pulsar timing
will be valuable for cross check.  From another perspective, coalescences of
compact objects provide tests based on transient events, 
while pulsar timing provides
tests based on long term observations. It will be interesting to
compare and combine the constraints from these two sets of experiments with
their different characteristics, probing different gravity regimes.

\vspace*{2mm} 
\Acknowledgements{\bahao We thank Renxin Xu for invitation and Weiwei Zhu for
reading the manuscript. This work has made use of NASA's Astrophysics Data
System.}


\end{multicols}

\begin{thebibliography}{200}

\bibitem{ein15}
A. Einstein,
in {\it Sitzungsberichte der K\"oniglich Preu\ss{}ischen Akademie
der Wissenschaften} (Berlin 1915), pp. 844--847.

\bibitem{ein15a}
A.~{Einstein},  
in {\it Sitzungsberichte der K{\"o}niglich Preu{\ss}ischen
Akademie der Wissenschaften} (Berlin 1915), pp. 831--839.

\bibitem{will14}
C.~M.~{Will},
{Living Rev. Relativ.} {\bf 17}, 4 (2014).

\bibitem{pdg14}
[Particle Data Group] 
K. A. Olive, K. Agashe, C. Amsler, M. Antonelli, J.-F.
Arguin, D. M. Asner, H. Baer, H. R. Band, R. M. Barnett, 
T. Basaglia, et al.,
Chin. Phys. C {\bf 38}, 090001 (2014).

\bibitem{he73}
S. W. Hawking and G. F. R. Ellis,
{\it The Large Scale Structure of Space-Time} (Cambridge
University Press, 1973).

\bibitem{ame13}
G. Amelino-Camelia,
{Living Rev. Relativ.} {\bf 16}, 5 (2013).

\bibitem{cfps12}
T. Clifton, P. G. Ferreira, A. Padilla, and C. Skordis,
Phys. Rept. {\bf 513}, 1 (2012).

\bibitem{goe12}
H.~{Goenner},
{Gen. Relat. Gravit.} {\bf 44}, 2077 (2012). 

\bibitem{jor55}
P.~{Jordan},
{\it Schwerkraft und Weltall}
(Vieweg, Braunschweig, 1955).

\bibitem{bek04}
J. D. Bekenstein,
{Phys. Rev. D} {\bf 70}, 083509 (2004).

\bibitem{bek04e}
J. D. Bekenstein,
{Phys. Rev. D} {\bf 71}, 069901 (2005).

\bibitem{hor09}
P. Ho{\v r}ava,
{Phys. Rev. D} {\bf 79}, 084008 (2009).

\bibitem{bps11}
D. Blas, O. Pujol{\`a}s, S. Sibiryakov,
{J. High Energy Phys.} {\bf 04}, 018 (2011).

\bibitem{bbc+15}
E. Berti, E. Barausse, V. Cardoso, L. Gualtieri, P. Pani, U. Sperhake, L. C. 
Stein, N. Wex, K. Yagi, T. Baker, et al.,
{Class.~Quantum~Grav.} {\bf 32}, 243001 (2015).

\bibitem{will93}
C. M. Will,
{\it Theory and Experiment in Gravitational Physics}
(Cambridge University Press, Cambridge, 1993).

\bibitem{dam09}
T.~{Damour}, 
In {\em Physics of Relativistic Objects in Compact Binaries: From
Birth to Coalescence}, edited by M.~{Colpi}, P.~{Casella}, V.~{Gorini},
U.~{Moschella}, and A.~{Possenti} (Astrophysics and Space Science Library,
Volume 359, 2009), pp. 1--41.

\bibitem{mtw73}
C.~W.~Misner, K.~S.~Thorne, and J.~A.~Wheeler, 
{\it Gravitation} (W.~H.~Freeman and Company, San Francisco, 1973).

\bibitem{dls15}
E.~{Di Casola}, S.~{Liberati}, and S.~{Sonego},
Am. J. Phys. {\bf 83}, 39 (2015).

\bibitem{sta03}
I. H. Stairs,
{Living Rev. Relativ.} {\bf 6}, 5 (2003).

\bibitem{wex14}
N. Wex, 
in {\it Brumberg Festschrift}, edited by S. M. Kopeikein (de Gruyter,
Berlin, 2014).

\bibitem{kos04}
V.~A.~Kosteleck{\'y},
{Phys. Rev. D} {\bf 69}, 105009 (2004).

\bibitem{ss09}
B. S. Sathyaprakash and B. F. Schutz,
Living Rev. Relativ. {\bf 12}, 2 (2009).

\bibitem{bs14}
A. Buonanno and B. S. Sathyaprakash,
in {\it General Relativity and Gravitation: A Centennial Perspective},
edited by A. Ashtekar, B. K. Berger, J. Isenberg,  and M. A. H.
MacCallum, in press, arXiv:1410.7832

\bibitem{nor68}
K.~{Nordtvedt},
{Phys.~Rev.} {\bf 170}, 1186 (1968).

\bibitem{mhb12}
J.~{M{\"u}ller}, F.~{Hofmann}, and L.~{Biskupek},
{Class.~Quantum~Grav.} {\bf 29}, 184006 (2012).

\bibitem{de93}
T. Damour and G.~Esposito-Far{\` e}se,
{Phys. Rev. Lett.} {\bf 70}, 2220 (1993).

\bibitem{de96a}
T. Damour and G.~Esposito-Far{\` e}se,
{Phys. Rev. D} {\bf 53}, 5541 (1996).

\bibitem{de92}
T. Damour and G.~Esposito-Far{\` e}se,
{Class. Quantum Grav.} {\bf 9}, 2093 (1992).

\bibitem{dam87}
T.~{Damour},
In {\em Three Hundred Years of Gravitation}, edited by S.W. Hawking and
W.~Israel (Cambridge University Press, Cambridge; New York, 1987), pp. 128--198.

\bibitem{lp01}
J.~M. {Lattimer} and M.~{Prakash},
{Astrophys.~J.} {\bf 550}, 426 (2001).

\bibitem{stob14}
M.~{Shibata}, K.~{Taniguchi}, H.~{Okawa}, and A.~{Buonanno},
{Phys.~Rev.~D} {\bf 89}, 084005 (2014).

\bibitem{ds91}
T.~{Damour} and G.~{Sch\"afer},
{Phys.~Rev.~Lett.} {\bf 66}, 2549 (1991).

\bibitem{sfl+05}
I. H. Stairs, A. J. Faulkner, A. G. Lyne, M. Kramer, D. R. Lorimer, M. A.
McLaughlin, R. N. Manchester, G. B. Hobbs, F. Camilo, A. Possenti, et al.,
{Astrophys. J.} {\bf 632}, 1060 (2005).

\bibitem{gsf+11}
M. E. Gonzalez, I. H. Stairs, R. D. Ferdman, P. C. C. Freire, D. J. Nice,
P. B. Demorest, S. M. Ransom, M. Kramer, F. Camilo, G. Hobbs, et al., 
{Astrophys. J.} {\bf 743}, 102 (2011).

\bibitem{fkw12}
P.~C.~C. {Freire}, M.~{Kramer}, and N.~{Wex},
{Class.~Quantum~Grav.} {\bf 29}, 184007 (2012).

\bibitem{rsa+14}
S. M. Ransom, I. H. Stairs, A. M. Archibald, J. W. T. Hessels, D. L.  Kaplan, M.
H. van Kerkwijk, J. Boyles,  A. T. Deller, S. Chatterjee, A.  Schechtman-Rook,
et al.,
{Nature} {\bf 505}, 520 (2014).

\bibitem{shao16}
L. Shao, 
{Phys. Rev. D} 93, 084023 (2016). 

\bibitem{fwe+12}
P. C. C. Freire, N. Wex, G. Esposito-Far{\` e}se, J. P.W. Verbiest, M. Bailes,
B. A. Jacoby, M. Kramer, I. H. Stairs, J.  Antoniadis, and G. H. Janssen,
{Mon. Not. R. Astron. Soc.} {\bf 423}, 3328 (2012).

\bibitem{kw09}
M.~{Kramer} and N.~{Wex},
{Class.~Quantum~Grav.} {\bf 26}, 073001 (2009).

\bibitem{mw13}
S.~{Mirshekari} and C.~M. {Will},
{Phys.~Rev.~D} {\bf 87}, 084070 (2013).

\bibitem{bbv08}
N.~D.~R.~{Bhat}, M.~{Bailes}, and J.~P.~W.~{Verbiest},
{Phys.~Rev.~D} {\bf 77}, 124017 (2008).

\bibitem{lwj+09}
K. Lazaridis, N. Wex, A. Jessner, M. Kramer, B. W. Stappers, G. H. Janssen, G.
Desvignes, M. B. Purver, I. Cognard, G. Theureau, et al.,
{Mon. Not. R. Astron. Soc.} {\bf 400}, 805 (2009).

\bibitem{afw+13}
J. Antoniadis, P. C. C. Freire, N. Wex, T. M. Tauris, R. S.  Lynch, M. H. van
Kerkwijk, M. Kramer, C. Bassa, V. S. Dhillon, T. Driebe, et al.,
{Science} {\bf 340}, 448 (2013).

\bibitem{dcl+16}
G.~{Desvignes}, R.~N.~{Caballero}, L.~{Lentati}, et al.,
{Mon.~Not.~R.~Astron.~Soc.}, {\bf 458}, 3341 (2016).

\bibitem{gw02}
J.-M. {G{\'e}rard} and Y.~{Wiaux},
{Phys.~Rev.~D} {\bf 66}, 024040 (2002).

\bibitem{ybyb14}
K. Yagi, D. Blas, E. Barausse, and N. Yunes,
{Phys. Rev. Lett.} {\bf 112}, 161101 (2014).

\bibitem{ybby14}
K. Yagi, D. Blas, E. Barausse, and N. Yunes,
{Phys. Rev. D} {\bf 89}, 084067 (2014).

\bibitem{ybby14e}
K. Yagi, D. Blas, E. Barausse, and N. Yunes,
{Phys. Rev. D} {\bf 90}, 069902 (2014).


\bibitem{lm07}
H. Lichtenegger and B. Mashhoon, in {\it The Measurement of Gravitomagnetism: A
Challenging Enterprise},  edited by L. Iorio (Nova Science, New York, 2007), pp.
13--25.

\bibitem{rin94}
W.~{Rindler},
{Phys.~Lett.~A} {\bf 187} 236 (1994).

\bibitem{fie56}
M.~{Fierz},
{Helv.~Phys.~Acta} {\bf 29} 128 (1956).

\bibitem{bd61}
C. Brans and R. H. Dicke,
{Phys. Rev.} {\bf 124}, 925 (1961).

\bibitem{bp95}
J. B. Barbour and H. Pfister (eds.),
{\it Mach's Principle: From Newton's Bucket to Quantum Gravity}
(Boston: Birkh\"auser, 1995).

\bibitem{bs97}
H. Bondi and J. Samuel,
{Phys. Lett. A} {\bf 228}, 121 (1997).

\bibitem{whi22}
A. N. Whitehead,
{\it The Principle of Relativity} (Cambridge: Cambridge University
Press, 1922).

\bibitem{sci53}
D. W. Sciama,
{Mon. Not. R. Astron. Soc.} {\bf 113}, 34 (1953).

\bibitem{will73}
C. M. Will,
{Astrophys. J.} {\bf 185}, 31 (1973).

\bibitem{rai75}
D. J. Raine,
{Mon. Not. R. Astron. Soc.} {\bf 171}, 507 (1975).

\bibitem{gw08}
G. Gibbons and C. M. Will,
{Stud. Hist. Phil. Sci.} {\bf 39}, 41 (2008).

\bibitem{cs58}
G. Cocconi and E. Salpeter,
{Nuovo Cimento} {\bf 10}, 646 (1958).

\bibitem{cs60}
G. Cocconi and E. Salpeter,
{Phys. Rev. Lett.} {\bf 4}, 176 (1960).

\bibitem{sw13}
L.~{Shao} and N.~{Wex},
{Class.~Quantum~Grav.} {\bf 30}, 165020 (2013).

\bibitem{swk15}
L. Shao, N. Wex, and M. Kramer, 
in {\it Proceedings of the Thirteenth Marcel Grossmann Meeting on General
Relativity}, edited by K. Rosquist, R. T. Jantzen, R. Ruffini (Singapore: World
Scientifici, 2015), pp. 1704--1706.

\bibitem{nor87}
K. Nordtvedt,
{Astrophys. J.} {\bf 320}, 871 (1987).

\bibitem{nw72}
K.~{Nordtvedt} and C.~M.~{Will},
{Astrophys. J.} {\bf 177}, 775 (1972).

\bibitem{wg76}
R. J. Warburton and J. M. Goodkind,
{Astrophys. J.} {\bf 208}, 881 (1976).

\bibitem{shi08}
S. Shiomi,
{Prog. Theor. Phys. Suppl.} {\bf 172}, 61 (2008).

\bibitem{lk04}
D. R. Lorimer and M. Kramer,
{\it Handbook of Pulsar Astronomy} (Cambridge: Cambridge
University Press, 2004).

\bibitem{aab+09}
B. A. Archinal, M. F. A'Hearn, E. Bowell, A. Conrad, G. J. Consolmagno, R.
Courtin, T. Fukushima, D. Hestroffer, J. L. Hilton, G. A. Krasinsky, et al.,
{Celest. Mech. Dyn. Astr.} {\bf 109}, 101 (2009).

\bibitem{ior14}
L. Iorio,
Mon. Not. R. Astron. Soc. {\bf 437}, 3482 (2014).

\bibitem{sku72}
A. Skumanich,
Astrophys. J. {\bf 171}, 565 (1972).

\bibitem{sck+13}
L.~{Shao}, R.~N. {Caballero}, M.~{Kramer}, N.~{Wex}, D.~J.~{Champion}, and
A.~{Jessner},
{Class.~Quantum~Grav.} {\bf 30}, 165019 (2013).

\bibitem{gjv+12}
L. Guillemot, T. J. Johnson, C. Venter, M. Kerr, B. Pancrazi, M. Livingstone, G.
H. Janssen, P. Jaroenjittichai, M. Kramer, I. Cognard, et al.,
{Astrophys. J.} {\bf 744}, 33 (2012).

\bibitem{ggr84}
J. Gil, P. Gronkowski, and W. Rudnicki,
{Astron. Astrophys.} {\bf 132}, 312 (1984).

\bibitem{jm01}
T. Jacobson and D. Mattingly,
{Phys. Rev. D} {\bf 64}, 024028 (2001).

\bibitem{bk06}
Q.~G.~Bailey and V.~A.~Kosteleck{\'y},
{Phys. Rev. D} {\bf 74}, 045001 (2006).

\bibitem{bl14}
D. Blas and E. Lim,
Int. J. Mod. Phys. D {\bf 23}, 1443009 (2014).

\bibitem{tas14}
J. D. Tasson,
{Rep. Prog. Phys.} {\bf 77}, 062901 (2014).

\bibitem{ks89a}
V.~A.~Kosteleck{\'y} and S.~Samuel,
{Phys. Rev. D} {\bf 39}, 683 (1989).

\bibitem{ks89b}
V.~A.~Kosteleck{\'y} and S.~Samuel,
{Phys. Rev. D} {\bf 40}, 1886 (1989).

\bibitem{gp99}
R. Gambini and J. Pullin,
{Phys. Rev.  D} {\bf 59}, 124021 (1999).

\bibitem{lk15}
J.~C.~Long and V.~A.~Kosteleck{\'y},
{Phys. Rev. D} {\bf 91}, 092003 (2015).

\bibitem{sb14}
A. R. Solomon and J. D. Barrow,
{Phys. Rev. D} {\bf 89}, 024001 (2014).

\bibitem{de92b}
T. Damour and G.~Esposito-Far{\` e}se,
{Phys. Rev. D} {\bf 46}, 4128 (1992).

\bibitem{lor08}
D. R. Lorimer,
{Living Rev. Relativ.} {\bf 11}, 8 (2008).

\bibitem{sw12}
L.~{Shao} and N.~{Wex},
{Class.~Quantum~Grav.} {\bf 29}, 215018 (2012).

\bibitem{shao14}
L.~{Shao},
{Phys. Rev. Lett.} {\bf 112}, 111103 (2014).

\bibitem{man15}
R. N. Manchester,
Int. J. Mod. Phys. D {\bf 24}, 1530018 (2015). 

\bibitem{wn72}
C.~M.~{Will} and K.~{Nordtvedt},
{Astrophys. J.} {\bf 177}, 757 (1972).

\bibitem{fm03}
Y.~{Fujii} and K.~I.~{Maeda}, {\em The ScalarÐTensor Theory of Gravitation}
(Cambridge: Cambridge University Press, 2003).

\bibitem{bcd96}
J. F. Bell, F. Camilo, and T. Damour,
{Astrophys. J.} {\bf 464}, 857 (1996).

\bibitem{wex00}
N. Wex,
in {\it IAU Colloq. 177: Pulsar Astronomy --- 2000 and Beyond}, edited by M.
Kramer, N. Wex and R. Wielebinski (Astronomical Society of the Pacific
Conference Series, Volume 202, 2000), pp.  113--116.

\bibitem{wk07}
N. Wex and M. Kramer,
{Mon. Not. R. Astron. Soc.} {\bf 380}, 455 (2007).

\bibitem{ksm+06}
M. Kramer, I. H. Stairs, R. N. Manchester, M. A. McLaughlin, A. G. Lyne, R. D.
Ferdman, M. Burgay, D. R. Lorimer, A. Possenti, N. D'Amico, et al.,
{Science} {\bf 314}, 97 (2006). 

\bibitem{dn93}
T. Damour and K. Nordtvedt,
{Phys. Rev. Lett.} {\bf 70}, 2217 (1993).

\bibitem{mwt08}
J. M\"uller, J. G. Williams, and S. G. Turyshev,
in {\it Lasers, Clocks and Drag-Free Control: Exploration of Relativistic
Gravity in Space}, edited by H. Dittus, C. L\"ammerzahl, and S. G. Turyshev
(Astrophysics and Space Science Library, Volume 349, 2008),
pp. 457--472.

\bibitem{bd96}
J. F. Bell and T. Damour,
{Class. Quantum Grav.} {\bf 13}, 3121 (1996).

\bibitem{zsd+15}
W. W. Zhu, I. H. Stairs, P. B. Demorest, D. J. Nice, J. A. Ellis, S. M. Ransom,
Z. Arzoumanian, K. Crowter, T. Dolch, R. D. Ferdman, et al.,
{Astrophys. J.} {\bf 809}, 41 (2015).

\bibitem{hil15}
D. Hilbert,
in {\it K\"onigliche Gesellschaft der Wissenschaften zu G\"ottingen.
Mathematisch-physikalische Klasse} (Nachrichten, 1916), pp. 395--407.

\bibitem{de96b}
T. Damour and G.~Esposito-Far{\` e}se,
{Phys. Rev. D} {\bf 54}, 1474 (1996).

\bibitem{de98}
T. Damour and G.~Esposito-Far{\` e}se,
{Phys. Rev. D} {\bf 58}, 042001 (1998).

\bibitem{ck97}
D.~Colladay and V.~A.~Kosteleck{\'y},
{Phys. Rev. D} {\bf 55}, 6760 (1997).

\bibitem{ck98}
D.~Colladay and V.~A.~Kosteleck{\'y},
{Phys. Rev. D} {\bf 58}, 116002 (1998).

\bibitem{hig14}
P.~W.~Higgs,
{Rev. Mod. Phys.} {\bf 86}, 851 (2014).

\bibitem{eng14}
F.~Englert,
{Rev. Mod. Phys.} {\bf 86}, 843 (2014).

\bibitem{bon15}
Y.~Bonder,
{Phys. Rev. D} {\bf 91}, 125002 (2015).

\bibitem{blu15}
R.~Bluhm,
{Phys. Rev. D} {\bf 91}, 065034 (2015).

\bibitem{nor76}
K.~Nordtvedt,
{Phys. Rev. D} {\bf 14}, 1511 (1976).

\bibitem{bkr15}
Q.~G.~Bailey, V.~A.~Kosteleck{\'y}, and R. Xu,
{Phys. Rev. D} {\bf 91}, 022006 (2015).

\bibitem{stt+15}
C.-G.~Shao, Y.-J.~Tan, W.-H.~Tan, S.-Q.~Yang, J.~Luo, and M.~E.~Tobar,
{Phys. Rev. D} {\bf 91}, 102007 (2015).

\bibitem{kt11}
V.~A.~Kosteleck{\'y} and J. D. Tasson,
{Phys. Rev. D} {\bf 83}, 016013 (2011).

\bibitem{lib13}
S. Liberati,
{Class. Quantum Grav.} {\bf 30}, 133001 (2013).

\bibitem{jty15}
R.~J.~Jennings, J.~D.~Tasson, and S.~Yang,
Phys. Rev. D {\bf 92}, 125028 (2015).

\bibitem{bcs07}
J. B. R. Battat, J. F. Chandler, and C. W. Stubbs,
{Phys. Rev. Lett.} {\bf 99}, 241103 (2007).

\bibitem{mch+08}
H. M\"uller, S.-W. Chiow, S. Herrmann, S. Chu, and K.-Y. Chung,
{Phys. Rev. Lett.} {\bf 100}, 031101 (2008).

\bibitem{cch+09}
K.-Y. Chung, S.-W. Chiow, S. Herrmann, S. Chu, and H. M\"uller,
{Phys. Rev. D} {\bf 80}, 016002 (2009).

\bibitem{beo13}
Q. G. Bailey, R. D. Everett, and J. M. Overduin,
{Phys. Rev. D} {\bf 80}, 102001 (2013).

\bibitem{hbp+15}
A. Hees, Q. G. Bailey, C. Le Poncin-Lafitte, A. Bourgoin, A. Rivoldini, B.
Lamine, F. Meynadier, C. Guerlin, and P. Wolf,
{Phys. Rev. D} {\bf 92}, 064049 (2015).

\bibitem{xie13}
Y.~Xie,
Res. Astron. Astrophys. {\bf 13}, 1 (2013).

\bibitem{shao14b}
L.~{Shao},
{Phys. Rev. D} {\bf 90}, 122009 (2014).

\bibitem{kr11}
V. A. Kosteleck{\'y} and N. Russell,
{Rev. Mod. Phys.} {\bf 83}, 11 (2011).

\bibitem{bcd+73}
P. L. Bender, D. G. Currie, S. K. Poultney, C. O. Alley, R. H. Dicke, D. T.
Wilkinson, D. H. Eckhardt, J. E. Faller, W. M. Kaula, J. D. Mulholland, et al.,
{Science} {\bf 182}, 229 (1973).

\bibitem{kc91}
M. Kasevich and S. Chu,
{Phys. Rev. Lett.} {\bf 67}, 181 (1991).

\bibitem{chu02}
S. Chu,
{Nature} {\bf 416}, 206 (2002).

\bibitem{edp+11}
C. W. F. Everitt, D. B. DeBra, B. W. Parkinson, J. P. Turneaure, J. W. Conklin,
M. I. Heifetz, G. M. Keiser, A. S. Silbergleit, T. Holmes, J. Kolodziejczak,
et al.,
{Phys. Rev. Lett.} {\bf 106}, 221101 (2011).

\bibitem{ove15}
J. M. Overduin,
{Class. Quantum Grav.} {\bf 32}, 224003 (2015).

\bibitem{dd86}
T.~Damour and N. Deruelle,
{Ann. Inst. Henri Poincar\'e A} {\bf 44}, 263 (1986).

\bibitem{dt92}
T.~Damour and J. H. Taylor,
{Phys. Rev. D} {\bf 45}, 1840 (1992).

\bibitem{hob13}
G. Hobbs,
{Class. Quantum Grav.} {\bf 30}, 224007 (2013).

\bibitem{mcl13}
M. A. McLaughlin,
{Class. Quantum Grav.} {\bf 30}, 224008 (2013).

\bibitem{kc13}
M. Kramer and D. J. Champion,
{Class. Quantum Grav.} {\bf 30}, 224009 (2013).

\bibitem{man13}
R. N. Manchester,
{Class. Quantum Grav.} {\bf 30}, 224010 (2013).

\bibitem{tll73}
K. S. Thorne, D. L. Lee, and A. P. Lightman,
{Phys. Rev. D} {\bf 7}, 3563 (1973).

\bibitem{will92}
C. M. Will,
{Astrophys. J.} {\bf 393}, 59 (1992).

\bibitem{bb86}
D. F.~Bartlett and D. van Buren,
{Phys. Rev. Lett.} {\bf 57}, 21 (1986).

\bibitem{will76}
C. M. Will,
{Astrophys. J.} {\bf 204}, 224 (1976).

\bibitem{uzan11}
J.-P. {Uzan},
{Living Rev.~Relativ.} {\bf 14}, 2 (2011).

\bibitem{wein72}
S.~{Weinberg},
{\it Gravitation and Cosmology: Principles and Applications of the General
Theory of Relativity} (John Wiley \& Sons, 1972).

\bibitem{PC13}
[Planck Collaboration] P. A. R. Ade, N. Aghanim, M. I. R. Alves, C.
Armitage-Caplan, M. Arnaud, M. Ashdown, F. Atrio-Barandela, J. Aumont, H.
Aussel, C. Baccigalupi, et al.,
{Astron. Astrophys.} {\bf 571}, 48 (2014).

\bibitem{dgt88}
T.~{Damour}, G.~W. {Gibbons}, and J.~H. {Taylor},
{Phys.~Rev.~Lett.} {\bf 61}, 1151 (1988).

\bibitem{hmb10}
F.~{Hofmann}, J.~{M{\"u}ller}, and L.~{Biskupek},
{Astron.~Astrophys.} {\bf 522}, L5 (2010).

\bibitem{fle+14}
A.~{Fienga}, J.~{Laskar}, P.~{Exertier}, H.~{Manche}, and M.~{Gastineau},
ArXiv preprint, arXiv:1409.4932 (2014).

\bibitem{nor90}
K.~{Nordtvedt},
{Phys.~Rev.~Lett.} {\bf 65}, 953 (1990).

\bibitem{ligo16}
The LIGO Scientific Collaboration and the Virgo Collaboration,
Phys. Rev. Lett. {\bf 116}, 061102 (2016).

\bibitem{ys13}
N. Yunes and X. Siemens,
{Living Rev. Relativ.} {\bf 16}, 9 (2013).

\bibitem{bppl13}
E.~{Barausse}, C.~{Palenzuela}, M.~{Ponce}, L.~{Lehner},
{Phys.~Rev.~D} {\bf 87}, 081506 (2013).

\end{thebibliography}
\end{document}